%% file: ms.tex
\pdfoutput=1
\documentclass[pageno]{jpaper}

\usepackage[normalem]{ulem}
\usepackage{mathptmx} % This is Times font
\usepackage{caption}
\newcommand{\ignore}[1]{}
\usepackage{multicol}
\usepackage{threeparttable}
\usepackage{fancyhdr}
\usepackage[normalem]{ulem}
\usepackage{microtype}
\usepackage[utf8]{inputenc}
\usepackage{amsmath}
\usepackage{amsfonts}
\usepackage{amssymb}
\usepackage{calrsfs}
\usepackage[mathscr]{euscript}
\usepackage{wasysym}
\usepackage{pifont}
\usepackage{graphics}
\usepackage{epsfig}
\usepackage{fancyhdr} 
\usepackage[utf8]{inputenc}
\usepackage[keeplastbox]{flushend}
\usepackage{balance}
\usepackage{subfloat}
\usepackage{mathtools}
\usepackage{booktabs, dcolumn}
\usepackage{graphicx} 
\usepackage{multirow}
\usepackage{subcaption}
\usepackage[table]{xcolor}

\usepackage{tikz}

\newcolumntype{d}[1]{D..{#1}}

\newcolumntype{C}[1]{>{\centering\let\newline\\\arraybackslash\hspace{1pt}}m{#1}}
\newcolumntype{L}[1]{>{\raggedleft\let\newline\\\arraybackslash\hspace{1pt}}m{#1}}

\begin{document}

\title{A Benchmark Suite for Evaluating Caches' Vulnerability to Timing Attacks}

\author{Shuwen Deng, Wenjie Xiong, Jakub Szefer\\ Yale University\\ \{shuwen.deng, wenjie.xiong, jakub.szefer\}@yale.edu}

\date{}
\maketitle

\thispagestyle{empty}

\begin{abstract}

\input{abstract}

\end{abstract}

\input{introduction}

\input{background}

\input{new_three_step}

\input{micro_benchmark}

\input{security_discussion}

\input{ctvs}

\input{validation}

\input{conclusion}

\clearpage

\bibliographystyle{plain}
\bibliography{references}

\end{document}

%% file: abstract.tex
Timing-based side or covert channels in processor caches continue to present a threat to computer systems, and they are the key to many of the recent Spectre and Meltdown attacks. Based on improvements to an existing three-step model for cache timing-based attacks, this work presents 88 Strong types of theoretical timing-based vulnerabilities in processor caches. To understand and evaluate all possible types of vulnerabilities in processor caches, this work further presents and implements a new benchmark suite which can be used to test to which types of cache timing-based attacks a given processor or cache design is vulnerable.
In total, there are 1094 automatically-generated test programs which cover the 88 theoretical vulnerabilities. The benchmark suite generates the Cache Timing Vulnerability Score which can be used to evaluate how vulnerable a specific cache implementation is to different attacks. A smaller Cache Timing Vulnerability Score means the design is more secure, and the scores among different machines can be easily compared. Evaluation is conducted on commodity Intel and AMD processors and shows the differences in processor implementations can result in different types of attacks that they are vulnerable to.
Beyond testing commodity processors, the benchmarks and the Cache Timing Vulnerability Score can be used to help designers of new secure processor caches evaluate their design's susceptibility to cache timing-based attacks.

%% file: introduction.tex
\section{Introduction}
\label{sec:introduction}

Cache timing channels have long been used to deploy attacks that can reconstruct 
sensitive data, especially secrets such as cryptographic keys~\cite{bonneau2006cache, bernstein2005cache, osvik2006cache}.
They usually make use of the timing difference in the memory operations between observing a cache hit and a cache miss
to derive the victim's secrets.
Since 2018, the cache timing-based attacks have gained new attention due to their use in 
Spectre~\cite{Kocher2018spectre} and Meltdown~\cite{Lipp2018meltdown}. 

With the long history of attacks, there are also many defenses proposed or deployed in software, e.g.,~\cite{crane2015thwarting, kong2009hardware}, 
and in hardware, 
e.g.,~\cite{zhang2012language,zhang2015hardware,wang2007new,wang2008novel,liu2014random,costan2016sanctum,wang2016secdcp,lee2005architecture,yan2017secure,domnitser2012non,liu2016catalyst,keramidas2008non,kayaalp2017ric,kirianskydawg,yan2018invisispec,bourgeat2018mi6,qureshi2018ceaser,wernerscattercache}.
However, existing
defenses can usually only prevent a subset of the attacks.
For example, RIC~\cite{kayaalp2017ric} cache has been shown to be able to defend the Prime+Probe type of attacks~\cite{osvik2006cache,percival2005cache},
but is vulnerable to the Flush+Reload~\cite{yarom2014flush+} type of attacks.
More importantly, so far most researchers have focused on coming up with individual attacks (and defenses for them), and there has been limited work on understanding 
all possible types of attacks.

To address the need to understand and evaluate all the different possible types of attacks, this 
paper presents both a theoretical model of all possible
timing-based attacks in caches, and a benchmark suite that can test for the theoretical vulnerabilities on real processors,
or simulations of new designs.
A key part of this work is an improved three-step model.
Compared to existing work~\cite{deng2019analysis}, the new three-step
model additionally considers: (1) differences in ``local'' and ``remote'' cores
for making the timing observations, and running the victim or the attacker code,
(2) the victim and attacker running in hyper-threading or time-slicing,
(3) using both read and write operations as memory accesses in the potential attacks (all but one prior work only considered reads), and
(4) two types of cache line invalidation operations, through flush instruction, or using cache coherence by writing on ``remote'' core to invalidate ``local'' core's cache lines.
The new three-step model  shows  32 new types of timing-based vulnerabilities not considered before,
which is in addition to new attacks found in the three-step model in~\cite{deng2019analysis}.

Based upon the new three-step model, we derive that
there are in total 88 {\em Strong} type of vulnerabilities, including 32 new ones.
For these vulnerabilities, we write scripts to automatically generate a total of 1094 benchmarks, which consider different variants for each vulnerability,
such as using reads vs. using writes.
The benchmarks are then used to test commodity Intel and AMD machines on both workstations and servers in lab and machines in Amazon's EC2 cloud.
The benchmarks can also be run on simulators to evaluate new types of secure processor caches. 
The benchmarks are further used to generate a new Cache Timing Vulnerability Score (CTVS),
which can help evaluate how different processors are vulnerable to different attacks -- differences of the processor implementations
mean that not all processors have the same vulnerabilities.
Based on the CTVS, designers can have a better understanding of the vulnerabilities in their designs and can develop new defenses.
The defenses can, for example, be customized to vulnerabilities that CTVS detects on the given processor, instead of a one-size-fits-all defense.

While the paper focuses on the 88 types of vulnerabilities for the benchmark design,
we also generated benchmarks for all  4913 possible three-step combinations in the three-step model.
Based on the results,
there are no more effective ones that are found apart from the vulnerabilities we derived.

\subsection{Contributions}

The contributions of this work are as follows:

\begin{itemize}

\item Derivation of a set of timing vulnerabilities on caches by the three-step model, 
based on improved modeling of the processor's behavior; 
a total of 88 {\em Strong} vulnerability types are
derived, including 32 new ones compared to prior work~\cite{deng2019analysis}.

\item Development of the first automatically-generated benchmark suite that is used to evaluate processor caches' vulnerability 
to all possible timing attacks. 

\item Evaluation of the benchmarks on Intel and AMD processors in the lab and on Amazon's EC2 cloud environment.

\item Generation of {\em Cache Timing Vulnerability Score} of processors 
to understand which attacks they are vulnerable to, evaluate
and customize defenses.

\item Validation of the three-step model using the benchmarks, 
demonstrating no benchmarks that give positive results beyond the ones corresponding to the 88 {\em Strong} types of vulnerabilities (in addition to {\em Weak} and repeat types).

\end{itemize}

\noindent The benchmarks and other code used in this paper will be released under open-source license at \url{https://urlredacted}.

%% file: background.tex
\section{Background}
\label{sec:background}

This section presents background on cache timing-based attacks,
the existing three-step model and existing metrics used to help understand vulnerabilities of caches to timing attacks.

\subsection{Timing of Memory Operations and Caches}

Processor caches are the key to helping processors maintain high performance when
accessing data.
However, the caches are of finite size, not all data can fit in them, and timing related to the memory 
operations involving caches, such as accesses resulting in cache hits and misses, 
can reveal information about the addresses or even data (for instruction caches it may be possible to reveal information about instructions as well).
In general, two types of memory-related operations exhibit timing variations that can be abused for 
timing-based side or covert channel attacks in processor data caches.
First, memory access operations, such as loads and stores can be fast (e.g., a cache hit) or slow (e.g., a cache miss).
Second, invalidation-related operations, such as cache flush, can be fast (e.g., there is no dirty data in cache so flush finishes quickly)
or slow (e.g., there is dirty data in the cache so it has to be written back, resulting in longer timing).

\subsection{Timing-Based Attacks on Processor Caches}
\label{subsec:software_attacks}

Researchers have proposed to use the timing differences in memory-related operations
to attack software, e.g.,~\cite{gullasch2011cache,percival2005cache,bernstein2005cache,bonneau2006cache,aciiccmez2006trace}.
Especially, the timing-based side-channel attacks often focus on cryptographic applications, e.g.,
attacks on software using AES with table lookups~\cite{daemen1999aes}.
Further, there are many timing-based covert-channel attacks where the sender and receiver cooperate 
to leak data, e.g., there are cache covert channels focusing on the last-level cache~\cite{liu2015last} and cross-core cache covert channels~\cite{maurice2015c5}. 
And most recently, timing-based channels are used as part of Spectre and Meltdown 
variants, e.g.,~\cite{Kocher2018spectre, Lipp2018meltdown, schwarz2018netspectre, koruyeh2018spectre}.

\subsection{Previous Three-Step Model for Timing-Based Attacks in Caches}

Recent work~\cite{deng2019analysis} has presented a
systematic approach to find all possible cache timing-based 
vulnerabilities.  
Their model is established based on two observations: 
all existing cache timing attacks focusing on the data are within three memory operations,
and timing attacks can be analyzed by checking the behavior of one cache block (since all blocks are updated in the same manner by the cache logic).

\begin{table}[!t]
\small
\caption{\small The 17 possible states for a single cache block of three-step model in \cite{deng2019analysis}.}
  \label{tbl:17_state}
  \fontsize{8}{9}\selectfont
  \begin{center}
    \begin{tabular}{|m{0.7cm}|m{7cm}| }
    \hline
     \centering \textbf{State} & \textbf{Description} \\ \hline\hline

     \centering $V_u$ & The cache block contains a specific memory location $u$ brought in by the victim, which is the address unknown to the attacker but within the set of sensitive memory locations $x$. 
     \\ \hline \centering $V_u^{inv}$ & The cache block state can be anything except $u$ in the cache block.
     The data and its address $u$ are ``removed'' from the cache block by the victim $V_u^{{inv}}$.
     \\ \hline\hline

     \centering $A_a$ \\or \\ $V_a$ & The cache block contains a specific memory location $a$, due to a memory access by the attacker, $A_a$,
     or the victim, $V_a$.
     The address $a$ is within the range of sensitive locations $x$ and known to the attacker.
     \\ \hline

     \centering $A_{a^{alias}}$\\ or\\ $V_{a^{alias}}$ & The cache block contains a memory address $a^{alias}$, due to a memory access by the attacker, $A_{a^{alias}}$,
     or the victim, $V_{a^{alias}}$.
     The address $a^{alias}$ is within the range $x$ and not the same as $a$,
     but it maps to the same cache block as $a$, i.e. it ``aliases'' to $a$.
     The address $a^{alias}$ is known to the attacker.
     \\ \hline

     \centering $A_{d}$  or  $V_{d}$ & The cache block contains a memory address $d$ due to a memory access by the attacker, $A_d$,
     or the victim, $V_d$.
     The address $d$ is not within the range $x$ and known to the attacker.
     \\ \hline

     \centering $A^{inv}$\\ or\\ $V^{inv}$ & The cache block is invalid.
     The data and its address are ``removed'' from the cache block by the attacker, $A^{inv}$, or the victim, $V^{inv}$,
     as a result of
     cache block being invalidated.
     \\ \hline

     \centering $A_a^{inv}$\\ or\\ $V_a^{inv}$ & The cache block state can be anything except $a$ in this cache block now.
     The data and its address $a$ are ``removed'' from the cache block by the attacker, $A_a^{inv}$, or the victim, $V_a^{inv}$.
     \\ \hline

     \centering $A_{a^{alias}}^{inv}$\\ or\\ $V_{a^{alias}}^{inv}$ & The cache block state can
     be anything except ${a^{alias}}$ in this cache block now.
     The data and its address ${a^{alias}}$ are ``removed'' from the cache block by the attacker, $A_{a^{alias}}^{inv}$, or the victim, $V_{a^{alias}}^{inv}$.
     \\ \hline

     \centering $A_d^{inv}$\\ or\\ $V_d^{inv}$ & The cache block state can be anything except $d$ in this cache block now.
     The data and its address $d$ are ``removed'' from the cache block by the attacker $A_d^{{inv}}$ or the victim $V_d^{{inv}}$.
     \\ \hline\hline

     \centering $\star$ & Any data, or no data, can be in the cache block.
     The attacker has no knowledge of the memory address in this cache block.
     \\ \hline

     \end{tabular}
  \end{center}
\end{table}

Following these observations, the work in~\cite{deng2019analysis} presented a three-step model focusing on one cache block 
for evaluating all possible timing-based attacks.  Further, a soundness analysis of the three-step model was performed to show that three steps
are sufficient to model all the timing-based attacks in caches.
In the three-step model, each step represents the state of the cache line after some memory-related operation is performed.
First, there is the initial step that sets the cache line into a known state.
Second, there is the following step that modifies the state of the cache line.
Finally, there is the last step, based on the timing of which, the change in the state of the cache line is observed.
Each of the steps can be performed by the attacker (A) or the victim (V).  The goal of~\cite{deng2019analysis}
is to find which three-step combinations can represent timing attacks from which the attacker learns
information about the unknown address accessed by the victim.
The authors list 17 possible states for a cache line, shown in Table~\ref{tbl:17_state}.
Among these states,
$V$ represents that the state is a result of the victim's operation, while $A$ represents that the state is a result of the attacker's operation.
$x$ denotes the set of virtual memory addresses storing addresses of sensitive data, and $u$ denotes the victim's secret address within~$x$ which is unknown to the attacker.
$a$, $a^{alias}$ and $d$ denote known memory addresses that map to the same cache line.
$d$ refers to an address outside of $x$, while the others are the address within~$x$.
The attacker's goal is to obtain $u$, which could be the same as $a$ or $a^{alias}$, maps to the same set as $a$, $a^{alias}$ and $d$, or not.

Given that there are 17 states and 3 steps, there are in total
$17^3=4913$ possible combinations of three steps that can be derived.
Their analysis demonstrated that 72 of the three-step combinations are 
{\em Strong} effective vulnerabilities, where the attacker can obtain the value of the unknown address $u$  unambiguously  with  fast or slow timing.
Meanwhile, 64 patterns were shown to be {\em Weak} effective vulnerabilities, where there are timing differences according to different values of $u$, 
but no single timing corresponds to a unique possible value of $u$.
With their analysis,~\cite{deng2019analysis} showed that 29 out of 72 vulnerabilities map to existing cache attacks.
\iffalse
, e.g.,  
$A^{{{inv}}} \rightsquigarrow V_u \rightsquigarrow V_a $(fast) maps to Cache Collision Attack~\cite{bonneau2006cache}, 
$A^{{{inv}}} \rightsquigarrow V_u \rightsquigarrow A_a$ (fast) maps to Flush + Reload Attack~\cite{yarom2014flush+},
$A_a \rightsquigarrow V_{u}^{inv} \rightsquigarrow A_a$ (slow) maps to SpectrePrime Attack~\cite{trippel2018meltdownprime},
$V_u \rightsquigarrow A_d \rightsquigarrow V_u$ (slow) maps to Evict + Time Attack~\cite{osvik2006cache},
$A_a \rightsquigarrow V_u \rightsquigarrow A_a$ (slow) maps to Prime + Probe Attack~\cite{osvik2006cache,percival2005cache},
$V_u \rightsquigarrow V_a \rightsquigarrow V_u$ (slow) maps to Bernstein's Attack~\cite{bernstein2005cache},
or $A_{a}^{inv} \rightsquigarrow V_u \rightsquigarrow V_{a}^{inv}$ (slow) maps to Flush + Flush Attack~\cite{gruss2016flush+}.
\fi
The other 43 types are new without corresponding attacks in literature at that time.

The existing work~\cite{deng2019analysis}, however, was limited in how they model the caches.  In 
particular, the work did not consider: read vs. write accesses, invalidation using flush 
instruction vs. invalidation using cache coherence, multithreading, and multicore system, 
and the possibility that there are more than just one ``fast'' and one ``slow'' timing in memory-related operation.
Also, they did not present any code or benchmarks for realizing and testing for the possible types
of vulnerabilities.  

Our work improves the model of caches, presents a set of 88 {\em Strong} vulnerability types, including 32 types not in prior work~\cite{deng2019analysis}, 
implements benchmarks for testing commodity processors, and validates the theoretical analysis by 
running all possible three-step combinations to show no other types of timing differences (and thus vulnerabilities) exist.

\subsection{Metrics for  Vulnerabilities in Caches}

Previously, \cite{zhang2014new} 
leveraged mutual information
to measure potential cache side-channel leakage. 
\cite{he2017secure} modeled cache interference using
probabilistic information flow graph. %,
However, \cite{zhang2014new} and \cite{he2017secure} 
only examined limited attacks including 
Evict + Time attack~\cite{osvik2006cache}, Cache Collision attack~\cite{bonneau2006cache}, 
Bernstein's attack~\cite{bernstein2005cache}, Prime + Probe attack~\cite{osvik2006cache,percival2005cache}, and Flush + Reload attack~\cite{yarom2014flush+}.
What's more, an analytical model was proposed in~\cite{domnitser2010predictive} to
track the fraction of the victim's
critical items accessible in the cache to determine leakage.
In a different work, SVF~\cite{demme2012side} metric measured information leakage by
measuring the
signal-to-noise ratio in an attacker's observations. 
CSV~\cite{zhang2013side} metric used direct correlation, in
place of phase correlation used by SVF, to measure leakage. 
The analytical model~\cite{domnitser2010predictive}, SVF~\cite{demme2012side} and CSV~\cite{zhang2013side} all only evaluated Prime + Probe attack~\cite{osvik2006cache,percival2005cache}.
Besides, the work in \cite{kopf2012automatic} quantified the cache side-channel leakage but mainly focused on access-driven attacks such as Evict + Time Attack~\cite{osvik2006cache}.
CacheD~\cite{wang2017cached} identified cache-based timing channels but mainly targeted 
 access-driven attacks such as Evict + Time Attack~\cite{osvik2006cache} 
and time-driven attacks such as Bernstein's Attack~\cite{bernstein2005cache}
 for timing-related attacks.
SCADET~\cite{sabbagh2018scadet} provided a side-channel detection tool targeting Prime + Probe attack~\cite{osvik2006cache,percival2005cache}.

This work is the first work to actually test for all possible timing-based vulnerabilities in caches, not just verify the cases concerning one or few attacks.
It is also the first that presents code which can be run on commodity processors
and which can generate the Cache Timing Vulnerability Score (CTVS) to evaluate different processors' caches.

%% file: new_three_step.tex
\iffalse
\begin{figure*}[t]
\centerline{\includegraphics[width=13cm]{figures/cache_hier.pdf}}
\caption{\small All possible data movements for 
timing observation step, i.e. $Step\;3$, in the three-step model.  Accesses causing the
data movement are read, write, or flush operations
with respect to a target cache line on the ``local'' core. 
\{1\} - \{13\}
correspond to local core's read operation to access clean L1 data, clean L2 data, clean L3 data, DRAM data, remote clean L1 data, remote clean L2 data, remote clean L3 data, dirty L1 data, dirty L2 data, dirty L3 data, remote dirty L1 data, remote dirty L2 data, or dirty L3 data, respectively;
\{14\} - \{26\}
correspond to local core's write operation to access clean L1 data, clean L2 data, clean L3 data, DRAM data, remote clean L1 data, remote clean L2 data, remote clean L3 data, dirty L1 data, dirty L2 data, dirty L3 data, remote dirty L1 data, remote dirty L2 data, or dirty L3 data, respectively;
\{27\} - \{39\}
correspond to local core's flush operation to access clean L1 data, clean L2 data, clean L3 data, DRAM data, remote clean L1 data, remote clean L2 data, remote clean L3 data, dirty L1 data, dirty L2 data, dirty L3 data, remote dirty L1 data, remote dirty L2 data, or dirty L3 data, respectively;
}
\label{fig:cache_hier}
\end{figure*}
\fi

\section{Modeling for Cache Timing Attacks}
\label{sec:new_three_step}

The goal of this work is to present the first set of benchmarks
which can be used to evaluate all the vulnerabilities of processor caches to
timing-based attacks.  Such attacks can be used, for example, 
by Spectre variants, e.g.,~\cite{Kocher2018spectre, Lipp2018meltdown, schwarz2018netspectre, koruyeh2018spectre},
to extract sensitive information.  With our benchmarks, it is possible
to help test processor caches or future secure cache designs,
and understand which timing attacks they are vulnerable to.

\subsection{Assumptions and Threat Model}
\label{subsec:ass_threat}
The goal of the benchmarks is to evaluate timing-based vulnerabilities in caches.
The presented benchmarks are not actual security exploits, rather they
implement memory-related operations that correspond to all possible timing-based attacks.
Each benchmark outputs whether there is a statistically significant timing difference
that the attacker could observe to extract information from the timing channel about the unknown address $u$ of the victim.

The current model focuses on all possible timing-based attacks in the L1 data cache.
The model includes uses of any memory-related operations (load, store, flush) and cache 
coherence protocol.
The model assumes a multi-core and possibly hyper-threading processor, 
with a cache hierarchy of 
local and remote L1 cache, L2 cache, and a shared L3 cache (which is possibly divided into different cache slices).

Current benchmarks do not consider timing-based attacks of other levels in cache hierarchy besides L1, but it should be straightforward to extend to the other levels.
We do not consider directory-related attacks~\cite{yan2019attack} or attacks based on replacement policy~\cite{xiong2019leaking}, 
but it should be possible to model these by adding more states to the model (and still keep an only total of three steps).
This work does not cover TLB attacks~\cite{gras2018translation, hund2013practical},
but there is already a theoretical model for TLBs~\cite{Deng:2019:ST:3307650.3322238}, and similar benchmarks can be developed
for TLB attacks (possibly merge with our benchmarks).

\begin{figure*}[t!]
    \centering

    \begin{subfigure}[t]{0.5\textwidth}
        \centering
        \includegraphics[width=9cm]{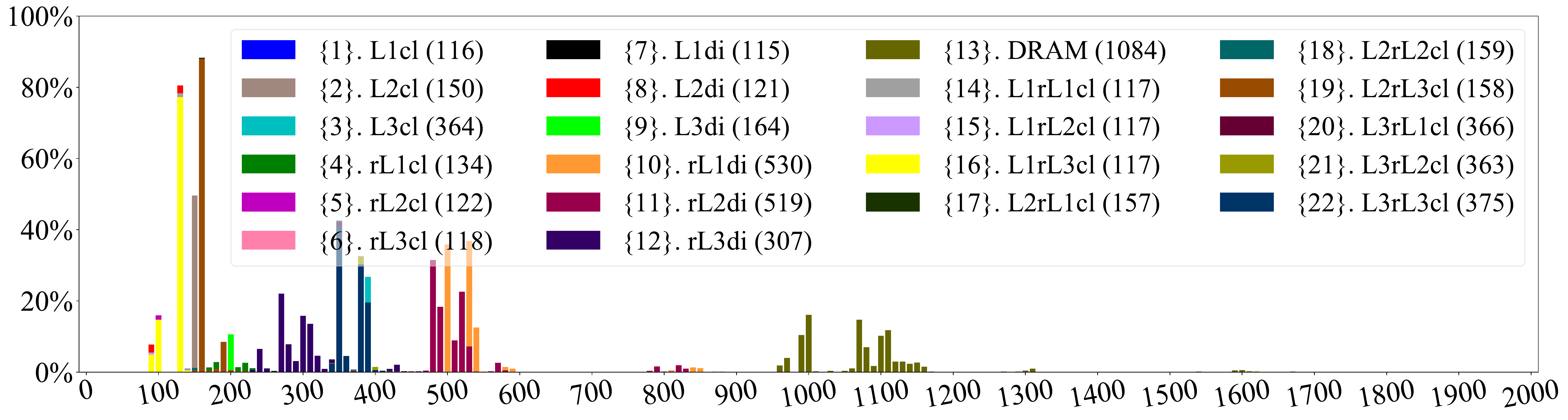}
        \caption{\small Timing of read access on Intel Xeon E5-1620 processor}
    \end{subfigure}%
    ~ 
    \begin{subfigure}[t]{0.5\textwidth}
        \centering
        \includegraphics[width=9cm]{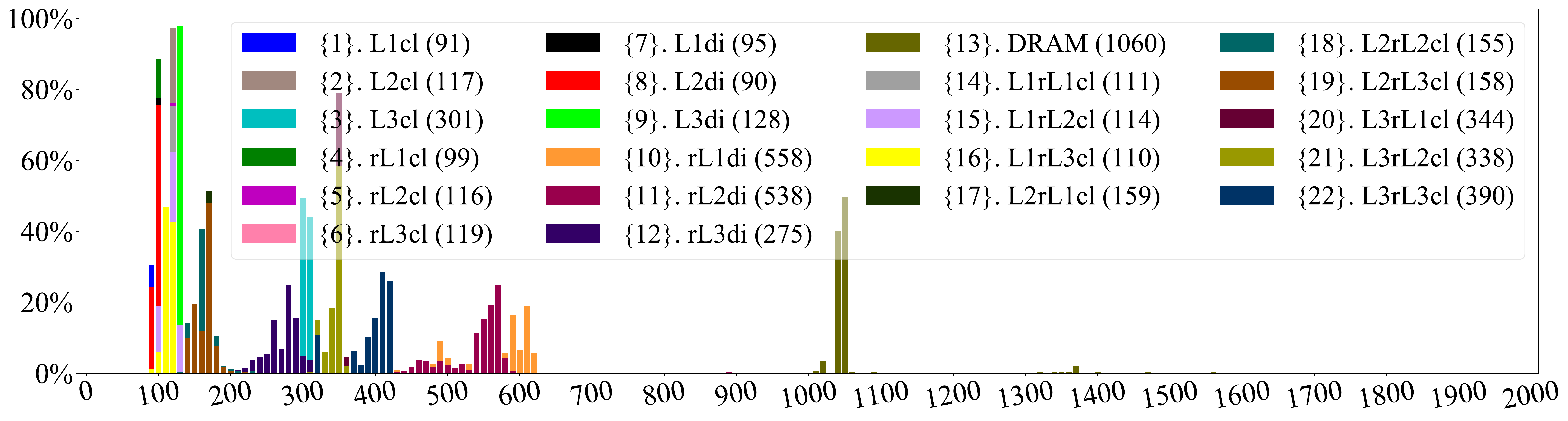}
        \caption{\small Timing of read access on Intel Xeon E5-2690 processor}
    \end{subfigure}

    \begin{subfigure}[t]{0.5\textwidth}
        \centering
        \includegraphics[width=9cm]{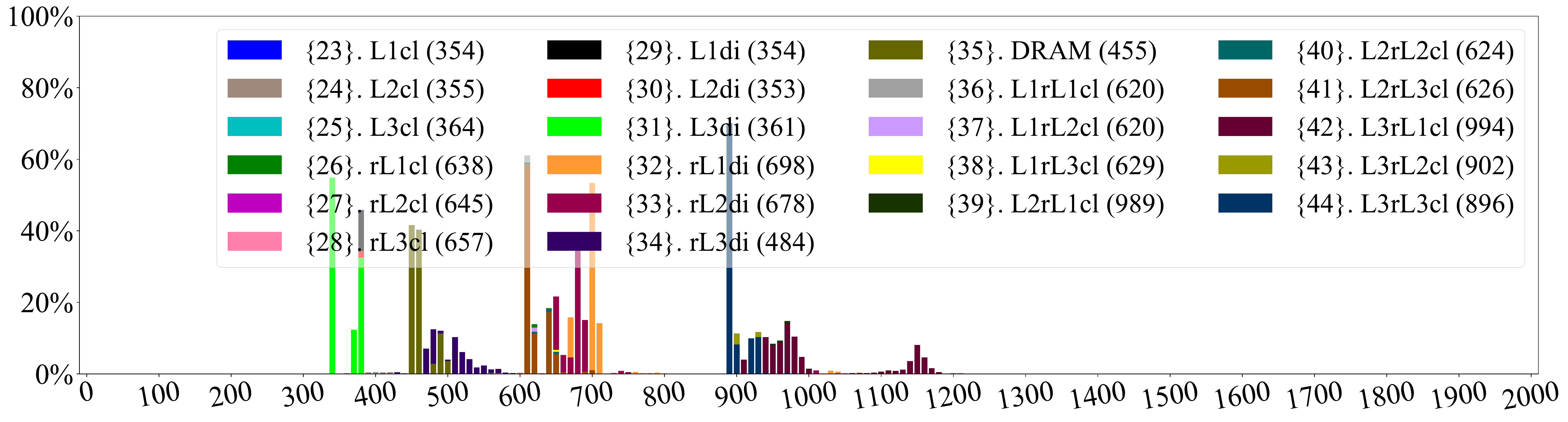}
        \caption{\small Timing of write access on Intel Xeon E5-1620 processor}
    \end{subfigure}%
    ~ 
    \begin{subfigure}[t]{0.5\textwidth}
        \centering
        \includegraphics[width=9cm]{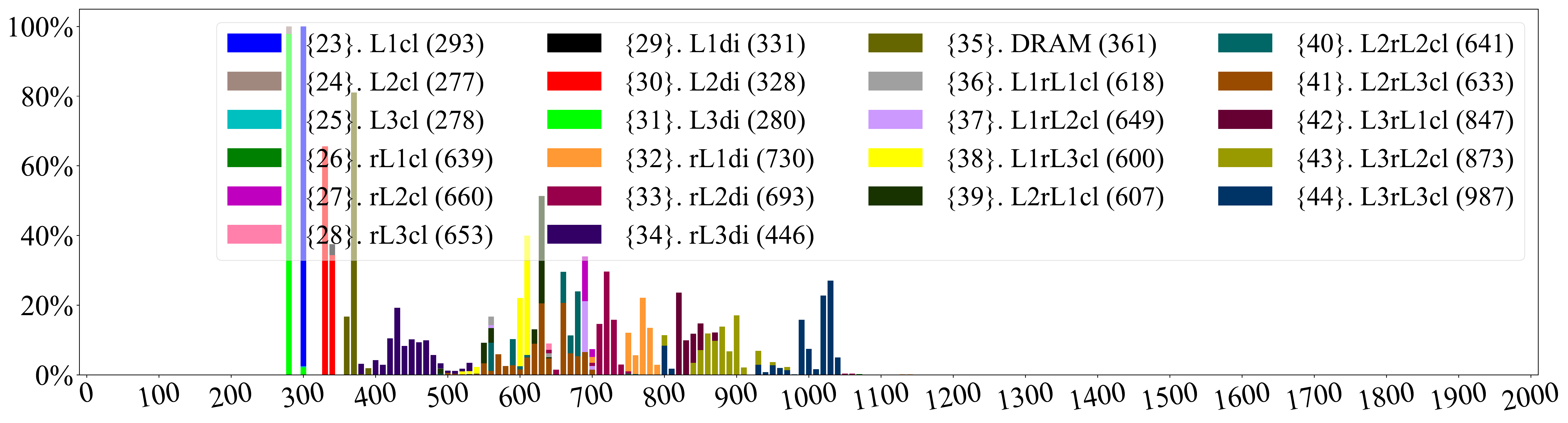}
        \caption{\small Timing of write access on Intel Xeon E5-2690 processor}
    \end{subfigure}

    \begin{subfigure}[t]{0.5\textwidth}
        \centering
        \includegraphics[width=9cm]{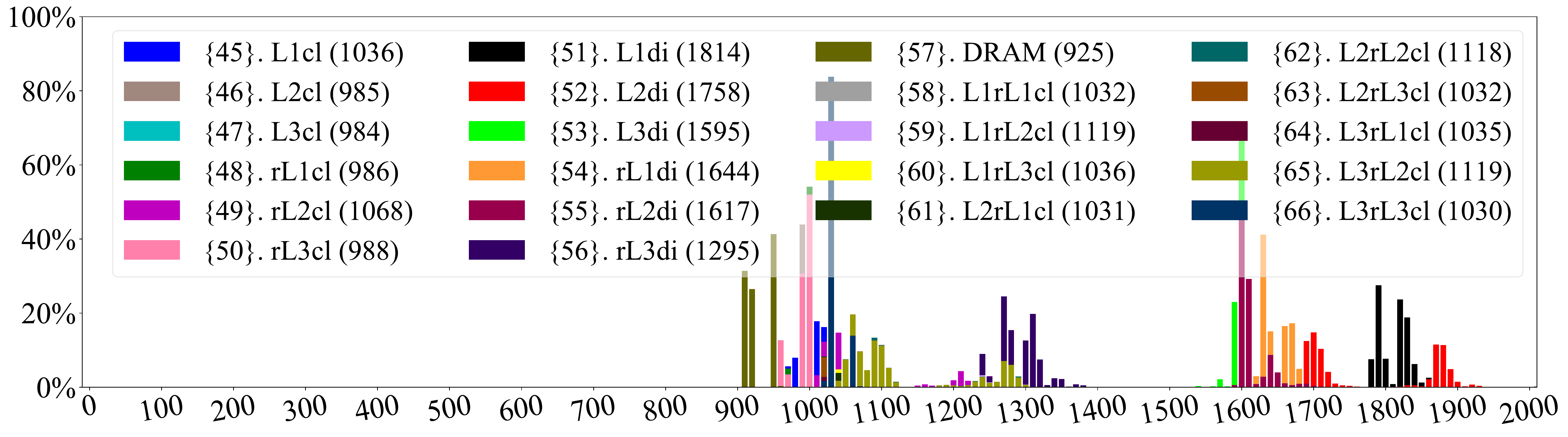}
        \caption{\small Timing of flush operation on  Intel Xeon E5-1620 processor}
    \end{subfigure}%
    ~ 
    \begin{subfigure}[t]{0.5\textwidth}
        \centering
        \includegraphics[width=9cm]{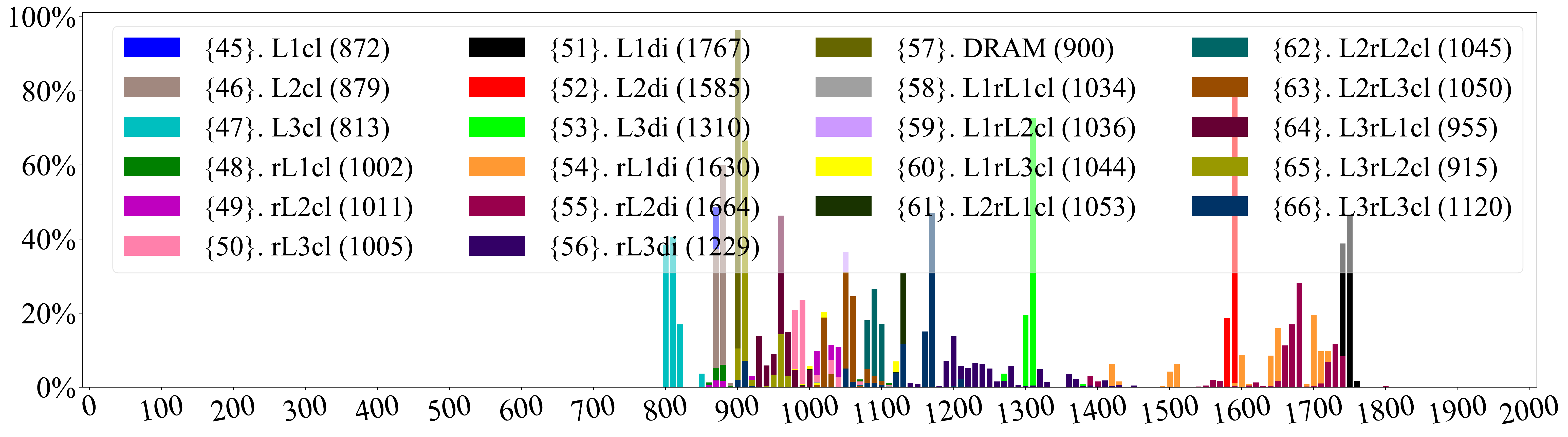}
        \caption{\small Timing of flush operation on Intel Xeon E5-2690 processor}
    \end{subfigure}

    \caption{\small 
    Histograms of read, write, and flush operations (each has 8 operations) timing under all possible data movements considered in this work.  
    The timing is for the timing observation step, i.e. $Step\;3$, in the tested three-step patterns.
    Note, different processors have different timing, and not all different types of data movements can be distinguished on different processors.
    The data is presented for Intel Xeon E5-1620 (a, c, e) and Intel Xeon E5-2690 (b, d, f) processors.
Numbers in the ``\{\}'' in the legend denote the different data movement types.
\{1\} - \{22\}
correspond to read operation, 
\{23\} - \{44\}
correspond to write operation,
\{45\} - \{66\}
correspond to flush operation,
 to access clean L1 data, clean L2 data, clean L3 data, remote clean L1 data, remote clean L2 data, remote clean L3 data, dirty L1 data, dirty L2 data, dirty L3 data, remote dirty L1 data, remote dirty L2 data, remote dirty L3 data, DRAM data, 
clean data in both L1 and remote L1, 
clean data in both L1 and remote L2, 
clean data in both L1 and remote L3, 
clean data in both L2 and remote L1, 
clean data in both L2 and remote L2, 
clean data in both L2 and remote L3, 
clean data in both L3 and remote L1, 
clean data in both L3 and remote L2, 
clean data in both L3 and remote L3, 
respectively.
    Numbers in the ``()'' in the legend show the average cycles needed for completing that type of memory operation.
    The $x$ axis shows the access latency in cycles.}
    \label{fig:hist}
\end{figure*}

The work considers more than just ``fast'' and ``slow'' timings
(as was not done by~\cite{deng2019analysis}).  This means that the influence of structures such as Miss Status Handling Registers (MSHRs),
load and store buffers between processor and caches, and line-fill buffers between cache levels are accounted for -- however,
benchmarks for timing attacks that are just due to these structures could likely be developed.

The flush operation here refers to the \textit{clflush} instruction in x86, %which 
which causes data to be flushed from all levels of caches back to main memory (including flushing data on other cores).
The timing measurements are from when each of the memory-related operations is issued until the instruction commits.

\subsection{Improved Modeling of Real Processors}
\label{sec:newthreestep}

We 
expand the model in~\cite{deng2019analysis} by considering more realistic cases for a processor's memory-related operation.

\textbf{Timing Observation on Local vs. Remote Core.}
Our cache attack model 
assumes a multi-core system and possibly a hyper-threading system as well.  
We model such a system using two cores: a ``local'' and ``remote'' core, each with L1, L2 and shared L3 caches.
The target cache block is located in the local core.
Remote core affects the target cache block on the 
local core by using cache coherence protocol. 
In particular, the remote core is used to perform write operations to invalidate the local core's data due to the cache coherence protocol.

For each read, write or flush operation, it may target the data that is in the local L1 cache, L2 cache, or L3 cache slice, 
or that is in the remote L1 cache, L2 cache, or L3 cache slice.
The cache block can be either in a clean or dirty state for the above 6 locations ($6*2=12$ types).
The clean data may also be in both local or remote core, which can be in any cache hierarchy (L1, L2, or L3 cache) for both cores ($3*3=9$ types).
Otherwise, the data is not in any level of the cache hierarchy, i.e., it is in the DRAM (1 type).
We consider all these 66 timings ($3$ operations $*\;(12+9+1)=66$) to be different from each other and use these 66 types of timings in our three-step cache simulator,
discusses in Section~\ref{subsec:improve_three},
to determine if a three-step combination can be used in an attack.

Figure~\ref{fig:hist} shows the histograms of these 66 types of timing observations for Intel Xeon E5-1620 and E5-2690 processors.
Based on the histograms, we found that some operations are differentiable from each other, while some are not. 
In general, the timing is processor-specific, so we need to consider and examine all 
various cache timings and cannot just assume ``fast'' and ``slow'' timings as was done by~\cite{deng2019analysis}.

\textbf{Hyper-Threading vs. Time-Slicing.}
Our cache model further considers that each core supports hyper-threading.  The victim and the attacker on one core can then either run in time-slicing setting
or run in parallel as two hyper-threads (if there is hyper-threading support in the processor).  For the case of accesses on 
``local'' vs. ``remote'' cores, the accesses on local and remote cores can be done in parallel.

\textbf{Read (Load) Access vs. Write (Store) Access.}
\label{subsubsec:read_write}
For memory-access-related operations, 
our model considers that they can  be either read (load) or write (store).
The timing of writes is not well explored in attacks, except for one work~\cite{storebypass2018}.
In this work, the model considers both reads and writes.
For example, for Flush + Reload attack, the previous attack~\cite{yarom2014flush+} uses the load operation in the final step to reload secret data and observe timing.
In  
our model, we also test store operation in the final step to access 
secret data
and reveal that attack with write in the final step is also possible, for example.

\textbf{Flush vs. Write Invalidation.}
In our model, 
we consider that a flush operation can be achieved by a $clflush$ type instruction, that flushes data from all caches back to main memory.  
Apart from that, 
our 
model considers the operation that can invalidate the local core's cache line by writing the corresponding 
line in the remote core, which will trigger cache coherence and result in the local cache 
line being invalidated.

\begin{table*}
\small
\begin{subtable}[t]{0.48\textwidth}
\centering
{\fontsize{8.5}{8.85}\selectfont%\small
\begin{tabular}[t]{|C{0.477cm}|C{0.54cm}|C{0.54cm}|C{0.54cm}|C{0.7cm}|C{1.42cm}|C{1.38cm}| }
    \hline

    \multirow{2}{*}{\shortstack{No.}} & \multicolumn{3}{ c |}{ Vulnerability Type} &  \multirow{2}{*}{\shortstack{Type}} & \multirow{2}{*}{\shortstack{Attack}}  & \multirow{2}{*}{\shortstack{Attack\\ Strategy}}\\ \cline{2-4}
    ~ & $S1$ & $S2$ & $S3$ & ~ & ~ & ~ \\ \hline\hline

    1 & $A^{{{inv}}}$ & $V_u$ & $V_a$  & $I$-$A$ & \cite{bonneau2006cache}&\multirow{4}{*}{\shortstack{Cache \\ Collision}} \\ \cline{1-6}
    2 & $V^{{{inv}}}$ & $V_u$ & $V_a$  & $I$-$A$ & \cite{bonneau2006cache} & ~ \\ \cline{1-6}
    3 & $A_a^{inv}$ & $V_{u}$ & $V_{a}$  & $I$-$A$ & \cite{bonneau2006cache} & ~\\ \cline{1-6}
    4 & $V_a^{inv}$ & $V_{u}$ & $V_{a}$  & $I$-$A$ & \cite{bonneau2006cache} & ~\\ \hline

    5 & $A_a^{inv}$ & $V_{u}$ & $A_{a}$  & $E$-$A$ & \cite{yarom2014flush+, gruss2015cache, schwarz2018netspectre} & \multirow{4}{*}{\shortstack{Flush\\ + Reload}}\\ \cline{1-6}
    6 & $V_a^{inv}$ & $V_{u}$ & $A_{a}$  & $E$-$A$ & \cite{yarom2014flush+, gruss2015cache, schwarz2018netspectre} & ~\\ \cline{1-6}
    7 & $A^{{{inv}}}$ & $V_u$ & $A_a$  & $E$-$A$ & \cite{yarom2014flush+, gruss2015cache, schwarz2018netspectre} & ~\\ \cline{1-6}
    8 & $V^{{{inv}}}$ & $V_u$ & $A_a$  & $E$-$A$ & \cite{yarom2014flush+, gruss2015cache, schwarz2018netspectre} & ~\\ \hline

    9 & $V_{u}^{inv}$ & $A_a$ & $V_u$  & $E$-$A$ & new in~\cite{deng2019analysis} & \multirow{2}{*}{\shortstack{Reload\\ + Time}}\\ \cline{1-6}
    10& $V_{u}^{inv}$ & $V_a$ & $V_u$  & $I$-$A$ & new in~\cite{deng2019analysis} & ~\\ \hline

    11 & $A_a$ & $V_{u}^{inv}$ & $A_a$  & $E$-$A$ & \cite{trippel2018meltdownprime} & \multirow{4}{*}{\shortstack{Flush\\ + Probe}}\\ \cline{1-6}
    12& $A_a$ & $V_{u}^{inv}$ & $V_a$  & $I$-$A$ & new in~\cite{deng2019analysis} & ~\\ \cline{1-6}
    13& $V_a$ & $V_{u}^{inv}$ & $A_a$  & $E$-$A$ & new in~\cite{deng2019analysis} & ~\\ \cline{1-6}
    14& $V_a$ & $V_{u}^{inv}$ & $V_a$  & $I$-$A$ & new in~\cite{deng2019analysis} & ~\\ \hline

    15 & $V_u$ & $A_{a}^{inv}$ & $V_u$  & $E$-$A$ & new in~\cite{deng2019analysis} & \multirow{2}{*}{\shortstack{Flush\\ + Time}}\\ \cline{1-6}
    16 & $V_u$ & $V_{a}^{inv}$ & $V_u$  & $I$-$A$ & new in~\cite{deng2019analysis} & ~\\ \hline

    \rowcolor{blue!12}
    17  & $A^{inv}$ & $V_{u}^{inv}$ & $A_a$  & $E$-$A$ & {\bf new} & ~\\ \cline{1-6}\rowcolor{blue!12}
    18& $A^{inv}$ & $V_{u}^{inv}$ & $V_a$  & $I$-$A$ & {\bf new} & ~\\ \cline{1-6}\rowcolor{blue!12}
    19& $V^{inv}$ & $V_{u}^{inv}$ & $A_a$  & $E$-$A$ & {\bf new} & ~\\ \cline{1-6}\rowcolor{blue!12}
    20& $V^{inv}$ & $V_{u}^{inv}$ & $V_a$  & $I$-$A$ & {\bf new} & \multirow{-4}{*}{\shortstack{Cache\\ Coherence\\ Flush\\ + Reload}}\\ \hline

    \rowcolor{blue!12}
    21 & $A_a^{inv}$ & $V_{u}^{inv}$ & $A_a$  & $E$-$SA$ & {\bf new} & ~\\ \cline{1-6}
    \rowcolor{blue!12}
    22& $A_a^{inv}$ & $V_{u}^{inv}$ & $V_a$  & $I$-$SA$ & {\bf new} & ~\\ \cline{1-6}
    \rowcolor{blue!12}
    23& $V_a^{inv}$ & $V_{u}^{inv}$ & $A_a$  & $E$-$SA$ & {\bf new} & ~\\ \cline{1-6}
    \rowcolor{blue!12}
    24& $V_a^{inv}$ & $V_{u}^{inv}$ & $V_a$  & $I$-$SA$ & {\bf new} & ~\\ \cline{1-6}
    \rowcolor{blue!12}
    25& $A_d^{inv}$ & $V_{u}^{inv}$ & $A_d$  & $E$-$S$ & {\bf new} & ~\\ \cline{1-6}
    \rowcolor{blue!12}
    26& $A_d^{inv}$ & $V_{u}^{inv}$ & $V_d$  & $I$-$S$ & {\bf new} & ~\\ \cline{1-6}
    \rowcolor{blue!12}
    27& $V_d^{inv}$ & $V_{u}^{inv}$ & $A_d$  & $E$-$S$ & {\bf new} & ~\\ \cline{1-6}
    \rowcolor{blue!12}
    28& $V_d^{inv}$ & $V_{u}^{inv}$ & $V_d$  & $I$-$S$ & {\bf new} & \multirow{-8}{*}{\shortstack{Cache\\ Coherence\\ Prime\\ + Probe}}\\ \hline

    \rowcolor{blue!12}
    29& $V_{u}^{inv}$ & $A_a^{inv}$ & $V_{u}$  & $E$-$SA$ & {\bf new} & ~\\ \cline{1-6}
    \rowcolor{blue!12}
    30& $V_{u}^{inv}$ & $V_a^{inv}$ & $V_{u}$  & $I$-$SA$ & {\bf new} & ~\\ \cline{1-6}
    \rowcolor{blue!12}
    31& $V_{u}^{inv}$ & $A_d^{inv}$ & $V_{u}$  & $E$-$S$ & {\bf new} & ~\\ \cline{1-6}
    \rowcolor{blue!12}
    32 & $V_{u}^{inv}$ & $V_d^{inv}$ & $V_{u}$  & $I$-$S$ & {\bf new} & \multirow{-4}{*}{\shortstack{Cache\\ Coherence\\ Evict\\ + Time}}\\ \hline
   
    33 & $V_u$ & $V_a$ & $V_u$  & $I$-$SA$ & \cite{bernstein2005cache} & \multirow{4}{*}{\shortstack{Bernstein's\\ Attack}}\\ \cline{1-6}
    34 & $V_u$ & $V_d$ & $V_u$  & $I$-$S$ & \cite{bernstein2005cache} & ~\\ \cline{1-6}
    35 & $V_d$ & $V_u$ & $V_d$  & $I$-$S$ & \cite{bernstein2005cache} & ~\\ \cline{1-6}
    36 & $V_a$ & $V_u$ & $V_a$  & $I$-$SA$ & \cite{bernstein2005cache} & ~\\ \hline
   
    37 & $V_d$ & $V_u$ & $A_d$  & $E$-$S$ & new in~\cite{deng2019analysis} & \multirow{2}{*}{\shortstack{Evict\\ + Probe}}\\ \cline{1-6}
    38 & $V_a$ & $V_u$ & $A_a$  & $E$-$SA$ & new in~\cite{deng2019analysis} & ~\\ \hline

    39 & $A_d$ & $V_u$ & $V_d$  & $I$-$S$ & new in~\cite{deng2019analysis} & \multirow{2}{*}{\shortstack{Prime\\ + Time}}\\ \cline{1-6}
    40 & $A_a$ & $V_u$ & $V_a$  & $I$-$SA$ & new in~\cite{deng2019analysis} & ~\\ \hline

    41 & $V_u$ & $A_d$ & $V_u$  & $E$-$S$ & \cite{osvik2006cache} & \multirow{2}{*}{\shortstack{Evict\\ + Time}}\\ \cline{1-6}
    42 & $V_u$ & $A_a$ & $V_u$  & $E$-$SA$ & \cite{osvik2006cache} & ~\\ \hline

    43 & $A_d$ & $V_u$ & $A_d$  & $E$-$S$ & \cite{osvik2006cache,percival2005cache,guanciale2016cache} & \multirow{2}{*}{\shortstack{Prime\\ + Probe}}\\ \cline{1-6}
    44 & $A_a$ & $V_u$ & $A_a$  & $E$-$SA$ & \cite{osvik2006cache,percival2005cache,guanciale2016cache} & ~\\ \hline

    \end{tabular}
    }

\caption{\small %footnotesize 
All the cache timing vulnerabilities with $S3$ as memory access operation.}
\label{tab:table1_a}
\end{subtable}
\hspace{\fill}
\begin{subtable}[t]{0.51\textwidth}
\small
\centering
{\fontsize{8.5}{8.2}\selectfont
\begin{tabular}[t]{|C{0.477cm}|C{0.545cm}|C{0.545cm}|C{0.545cm}|C{0.7cm}|C{1.405cm}|C{1.66cm}|}
    \hline

      \multirow{2}{*}{\shortstack{No.}} & \multicolumn{3}{ c |}{ Vulnerability Type} &  \multirow{2}{*}{\shortstack{Type}} & \multirow{2}{*}{\shortstack{Attack}}  & \multirow{2}{*}{\shortstack{Attack\\ Strategy}}\\ \cline{2-4}
    ~ & $S1$ & $S2$ & $S3$ & ~ & ~ & ~ \\ \hline\hline

    45& $A^{inv}$ & $V_{u}$ & $V_a^{inv}$  & $I$-$A$ & new in~\cite{deng2019analysis}  & \multirow{2}{*}{\shortstack{Cache \\Collision Inv.}} \\ \cline{1-6}
    46& $V^{inv}$ & $V_{u}$ & $V_a^{inv}$  & $I$-$A$ & new in~\cite{deng2019analysis}  & ~ \\ \hline

    47 & $A_{a}^{inv}$ & $V_u$ & $V_{a}^{inv}$  & $I$-$A$ & \cite{gruss2016flush+} & \multirow{4}{*}{\shortstack{Flush +\\ Flush}} \\ \cline{1-6}
    48& $V_{a}^{inv}$ & $V_u$ & $V_{a}^{inv}$  & $I$-$A$ & \cite{gruss2016flush+}  & ~ \\ \cline{1-6}
    49& $A_{a}^{inv}$ & $V_u$ & $A_{a}^{inv}$  & $E$-$A$ & \cite{gruss2016flush+}  & ~ \\ \cline{1-6}
    50& $V_{a}^{inv}$ & $V_u$ & $A_{a}^{inv}$  & $E$-$A$ & \cite{gruss2016flush+}  & ~ \\ \hline

    51 & $A^{inv}$ & $V_{u}$ & $A_a^{inv}$  & $E$-$A$ & new in~\cite{deng2019analysis}  & \multirow{2}{*}{\shortstack{Flush + \\Reload  Inv.}} \\ \cline{1-6}
    52& $V^{inv}$ & $V_{u}$ & $A_a^{inv}$  & $E$-$A$ & new in~\cite{deng2019analysis}  & ~ \\ \hline

    53 & $V_{u}^{inv}$ & $A_a$ & $V_u^{inv}$  & $E$-$A$ & new in~\cite{deng2019analysis}  & \multirow{2}{*}{\shortstack{Reload + \\Time  Inv.}}\\ \cline{1-6}
    54& $V_{u}^{inv}$ & $V_a$ & $V_u^{inv}$  & $I$-$A$ & new in~\cite{deng2019analysis}  & ~ \\ \hline

    55 & $A_a$ & $V_{u}^{inv}$ & $A_a^{inv}$  & $E$-$A$ & new in~\cite{deng2019analysis}  & \multirow{4}{*}{\shortstack{Flush + \\Probe Inv.}} \\ \cline{1-6}
    56& $A_a$ & $V_{u}^{inv}$ & $V_a^{inv}$  & $I$-$A$ & new in~\cite{deng2019analysis}  & ~ \\ \cline{1-6}
    57& $V_a$ & $V_{u}^{inv}$ & $A_a^{inv}$  & $E$-$A$ & new in~\cite{deng2019analysis}  & ~ \\ \cline{1-6}
    58& $V_a$ & $V_{u}^{inv}$ & $V_a^{inv}$  & $I$-$A$ & new in~\cite{deng2019analysis}  & ~ \\ \hline

    59 & $V_u$ & $A_{a}^{inv}$ & $V_u^{inv}$  & $E$-$A$ & new in~\cite{deng2019analysis}  & \multirow{2}{*}{\shortstack{Flush + \\Time  Inv.}} \\ \cline{1-6}
    60& $V_u$ & $V_{a}^{inv}$ & $V_u^{inv}$  & $I$-$A$ & new in~\cite{deng2019analysis}  & ~ \\ \hline

    \rowcolor{blue!12}
     61& $A^{inv}$ & $V_{u}^{inv}$ & $A_a^{inv}$  & $E$-$A$ & {\bf new}  & ~ \\ \cline{1-6}
     \rowcolor{blue!12}
    62& $A^{inv}$ & $V_{u}^{inv}$ & $V_a^{inv}$  & $I$-$A$ & {\bf new}  & ~ \\ \cline{1-6}
    \rowcolor{blue!12}
    63& $V^{inv}$ & $V_{u}^{inv}$ & $A_a^{inv}$  & $E$-$A$ & {\bf new}  & ~ \\ \cline{1-6}
    \rowcolor{blue!12}
   64 & $V^{inv}$ & $V_{u}^{inv}$ & $V_a^{inv}$  & $I$-$A$ & {\bf new}  & \multirow{-4}{*}{\shortstack{Cache\\ Coherence\\  Flush +\\ Reload Inv.}} \\ \hline

  \rowcolor{blue!12}
  65 & $A_a^{inv}$ & $V_{u}^{inv}$ & $A_a^{inv}$  & $E$-$SA$ & {\bf new}  & ~ \\ \cline{1-6}
  \rowcolor{blue!12}
  66& $A_a^{inv}$ & $V_{u}^{inv}$ & $V_a^{inv}$  & $I$-$SA$ & {\bf new}  & ~ \\ \cline{1-6}
  \rowcolor{blue!12}
  67& $V_a^{inv}$ & $V_{u}^{inv}$ & $A_a^{inv}$  & $E$-$SA$ & {\bf new}  & ~ \\ \cline{1-6}
  \rowcolor{blue!12}
  68& $V_a^{inv}$ & $V_{u}^{inv}$ & $V_a^{inv}$  & $I$-$SA$ & {\bf new}  & ~ \\ \cline{1-6}
  \rowcolor{blue!12}
  69 & $A_d^{inv}$ & $V_{u}^{inv}$ & $A_d^{inv}$  & $E$-$S$ & {\bf new}  & ~ \\ \cline{1-6}
   \rowcolor{blue!12}
  70& $A_d^{inv}$ & $V_{u}^{inv}$ & $V_d^{inv}$  & $I$-$S$ & {\bf new}  & ~ \\ \cline{1-6}
  \rowcolor{blue!12}
  71& $V_d^{inv}$ & $V_{u}^{inv}$ & $A_d^{inv}$  & $E$-$S$ & {\bf new}  & ~ \\ \cline{1-6}
  \rowcolor{blue!12}
  72& $V_d^{inv}$ & $V_{u}^{inv}$ & $V_d^{inv}$  & $I$-$S$ & {\bf new}  & \multirow{-8}{*}{\shortstack{Cache\\ Coherence\\ Prime +\\ Probe Inv.}} \\ \hline

  \rowcolor{blue!12}
  73 & $V_{u}^{inv}$ & $A_a^{inv}$ & $V_{u}^{inv}$  & $E$-$SA$ & {\bf new}  & ~ \\ \cline{1-6}
  \rowcolor{blue!12}
  74& $V_{u}^{inv}$ & $V_a^{inv}$ & $V_{u}^{inv}$  & $I$-$SA$ & {\bf new}  & ~ \\ \cline{1-6}
  \rowcolor{blue!12}
  75& $V_{u}^{inv}$ & $A_d^{inv}$ & $V_{u}^{inv}$  & $E$-$S$ & {\bf new}  & ~ \\ \cline{1-6}
  \rowcolor{blue!12}
  76& $V_{u}^{inv}$ & $V_d^{inv}$ & $V_{u}^{inv}$  & $I$-$S$ & {\bf new}   & \multirow{-4}{*}{\shortstack{Cache\\ Coherence\\ Evict + \\Time Inv.}} \\ \hline

    77 & $V_u$ & $V_{a}$ & $V_u^{inv}$  & $I$-$SA$ & new in~\cite{deng2019analysis}  & \multirow{4}{*}{\shortstack{Bernstein's \\ Inv.\\ Attack}} \\ \cline{1-6}
    78& $V_u$ & $V_{d}$ & $V_u^{inv}$  & $I$-$S$ & new in~\cite{deng2019analysis}  & ~ \\ \cline{1-6}
    79 & $V_d$ & $V_{u}$ & $V_d^{inv}$  & $I$-$S$ & new in~\cite{deng2019analysis}  & ~ \\ \cline{1-6}
    80 & $V_a$ & $V_u$ & $V_{a}^{inv}$  & $I$-$SA$ & new in~\cite{deng2019analysis}  & ~ \\ \hline

    81 & $V_d$ & $V_{u}$ & $A_d^{inv}$  & $E$-$S$ & new in~\cite{deng2019analysis}  & \multirow{2}{*}{\shortstack{Evict +\\ Probe Inv.}} \\ \cline{1-6}
    82 & $V_a$ & $V_u$ & $A_{a}^{inv}$  & $E$-$SA$ & new in~\cite{deng2019analysis}  & ~ \\ \hline

    83 & $A_d$ & $V_{u}$ & $V_d^{inv}$  & $I$-$S$ & new in~\cite{deng2019analysis}  & \multirow{2}{*}{\shortstack{Prime +\\ Time Inv.}} \\ \cline{1-6}
    84 & $A_a$ & $V_u$ & $V_{a}^{inv}$  & $I$-$SA$ & new in~\cite{deng2019analysis}  & ~ \\ \hline

    85   & $V_u$ & $A_{d}$ & $V_u^{inv}$  & $E$-$S$ & new in~\cite{deng2019analysis}  & \multirow{2}{*}{\shortstack{Evict +\\ Time Inv.}} \\ \cline{1-6}
    86 & $V_u$ & $A_{a}$ & $V_u^{inv}$  & $E$-$SA$ & new in~\cite{deng2019analysis}  & ~ \\ \hline

    87 & $A_d$ & $V_{u}$ & $A_d^{inv}$  & $E$-$S$ & new in~\cite{deng2019analysis}  & \multirow{2}{*}{\shortstack{Prime +\\ Probe Inv.}} \\ \cline{1-6}
    88 & $A_a$ & $V_u$ & $A_{a}^{inv}$  & $E$-$SA$ & new in~\cite{deng2019analysis}  & ~ \\ \hline

    \end{tabular} 
   }

\caption{\small 
All the cache timing vulnerabilities with $S3$ as invalidation operation.}
\label{tab:table1_b}
\end{subtable}

\caption{\small 
The table shows all the L1 cache timing-based vulnerabilities.
The {\em No.} column assigns each type of vulnerability a number.
The {\em Vulnerability Type} column shows the three steps that define each vulnerability. $S\#$ denotes $Step\#$.
The {\em Type} column proposes the categorization the vulnerability belongs to. 
``$E$'' and ``$I$'' are for internal and external interference types, respectively.
 ``$S$'', ``$A$'' and ``$SA$'' are set-based, address-based types
 and the types that are both set-based and address-based, respectively.
The {\em Attack} column shows if a vulnerability has been
previously presented in the literature.
The {\em Attack Strategy} column gives a common name for each set of
vulnerabilities that would be exploited in an attack in a similar manner.
Inv. means invalidation.
Light-blue colored rows are the vulnerabilities which are first presented in this work.
}
\label{tbl:attack_list_all}
\end{table*}

\section{Derivation of All Vulnerabilities}
\label{subsec:improve_three}

In this work, we build a new cache three-step simulator based on the new model discussed in Section~\ref{sec:newthreestep}.
It considers different memory-related operations and differentiates 
among the 66 timing variations discussed in Section~\ref{sec:newthreestep}
that are related to L1 cache timing-based attack for the final timing observation 
step.
Further, we give categorizations of vulnerabilities to find common features that attacks exploit.

\subsection{Judging the Effectiveness of Three-Step Combination}
\label{subsec:attackers_analysis}

In order for a three-step combination to be effective for an attack, 
at least the unknown victim's address $u$ should be involved in one of the three steps since $u$ is the unknown secret the attacker tries to learn.
In this case, the vulnerability will have $V_u$ or $V_u^{inv}$ as one or more of the three-steps to represent the
operations on the secret $u$.

Based on the 17 states shown in Table~\ref{tbl:17_state}, for the three-step model,
the attacker tries to learn the value of $u$ by guessing if $u$ equals: $a$, $a^{alias}$ or $NIB$.  
$a$ denotes the address that is within the set of sensitive locations $x$ and maps to the target cache line.
$a^{alias}$ denotes any data address that belongs to sensitive locations $x$ and also maps to the cache line but is not $a$.
Apart from all possible sensitive address mapping to the target cache line,
$u$ can possibly not map to the target cache line the attacker is measuring. 
We denote these addresses as ${NIB}$ (not-in-block).
Therefore,
$u$ can be either $a$, $a^{alias}$, or ${NIB}$.
If the attacker is able to unambiguously correlate the timing to one of the three values, he or she
is able to learn the value of $u$ and the corresponding three-steps is a {\em Strong} type vulnerability.  
Meanwhile, if the attacker is not able to clearly distinguish whether
$u$ is $a$, $a^{alias}$, or ${NIB}$ based on the timing, 
but there are still timing differences observed,
then the corresponding attacks belong to {\em Weak} type of vulnerabilities.
Otherwise, if the timing is always the same regardless of different values of $u$, it will be an {\em Ineffective} three-step combination.

\begin{figure}[t]
\centerline{\includegraphics[width=9cm]{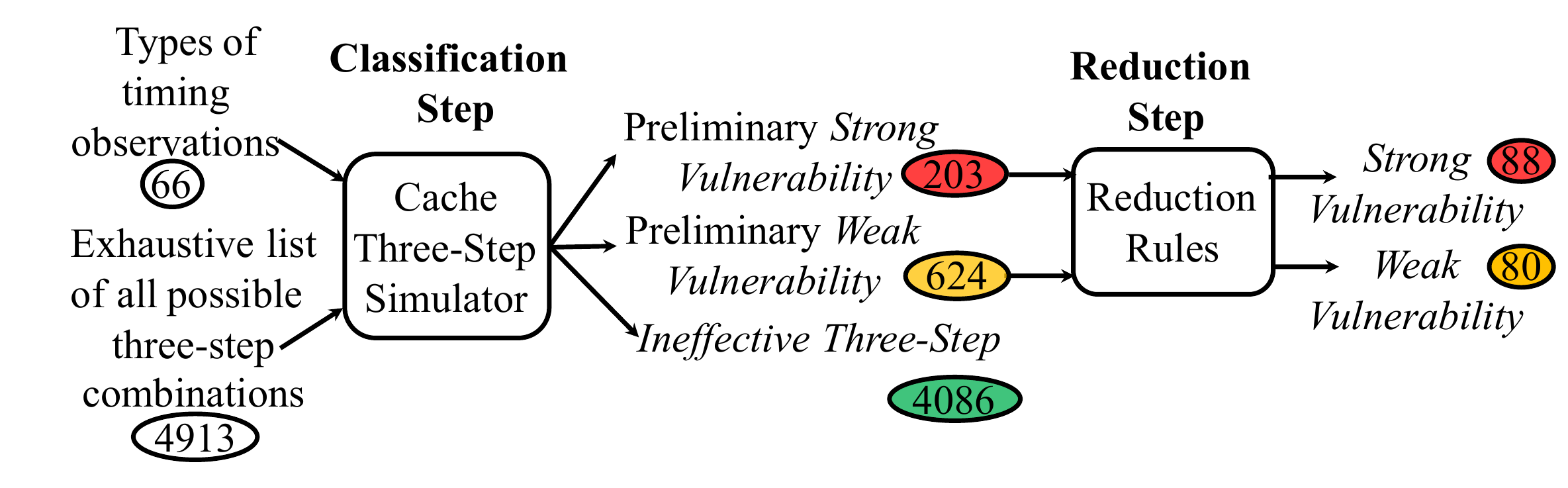}}
\caption{\small The derivation process of all the {\em Strong} and {\em Weak} types of L1 cache timing-based vulnerabilities.}
\label{fig:vul_derivation}
\end{figure}

\subsection{New Cache Three-Step Simulator}
Figure~\ref{fig:vul_derivation} shows the derivation process of vulnerabilities. 
We write Python scripts to develop the cache three-step simulator.
The simulator takes all 4913 combinations and 66 types of timing observations 
as input, 
checks 
and outputs the three-steps that 
belong to {\em Strong}, {\em Weak} and {\em Ineffective} types, respectively.
For the step that is
$V_u$, the simulator checks the timing when $V_u$ is $V_a$, $V_{a^{alias}}$, and $V_{NIB}$.
For the step that is
$V_u^{inv}$, the simulator checks the timing when $V_u^{inv}$ is $V_a^{inv}$, $V_{a^{alias}}^{inv}$, and $V_{NIB}^{inv}$.
The timing variance exists if different possible values of $u$ correspond 
to different timings of the 66 types.
We enumerate all possible operations (read/write for access, remote write/flush for invalidation) for a step and consider different timings for each operation.
Therefore, each three-step pattern may have several types of timing observations.
While in~\cite{deng2019analysis}, 
only two timing differences (``fast'' and ``slow'') were considered when comparing the timing variation to identify effective vulnerabilities.
The rules from~\cite{deng2019analysis} to remove repeat and redundant three-step patterns are used for our work.

As shown in Figure~\ref{fig:vul_derivation},
based on the much finer-grained categorization of timing differences, 
we derived in total 88 {\em Strong} effective vulnerabilities and 80 {\em Weak} effective vulnerabilities after removing repeat three-step patterns.
They are shown in Table~\ref{tbl:attack_list_all}, 
where light-blue colored rows (in total 32 types) are the new vulnerabilities (compared to~\cite{deng2019analysis}) 
which we  found through 
running of new cache three-step simulator (16 types of the original {\em Strong} effective vulnerabilities become {\em Weak} vulnerabilities when considering multi-core systems).
We provide new names for the new attacks in {\em Attack Strategy} in Table~\ref{tbl:attack_list_all} while re-use existing names if the attacks were presented before.
As verified in Section~\ref{sec:validation} through the tested processors, there are no other effective vulnerabilities 
except the types
we derive in Table~\ref{tbl:attack_list_all}, further strengthening our findings.

\subsection{Categorizations of the Vulnerabilities}
We first categorize different vulnerabilities as based on internal ($I$) or external ($E$) interference.
The types that
only involve the victim's behavior, $V$, in the states of $Step\;2$ and $Step\;3$ are internal interference vulnerabilities ($I$).
The remaining ones are external interference ($E$) vulnerabilities.

In~\cite{deng2019analysis, zhang2014new}, the vulnerabilities are categorized as hit-based and miss-based 
vulnerabilities, based on the cache behaviors the
attackers want to observe (cache misses or hits).
This definition does not fit our model since there are 
different types of timings for L1 data hits and misses in the real machines.
For example, attacks can derive information on accesses using timing difference from two types of cache misses, which are not covered by the original categorizations.

Therefore, we create new categorizations for the vulnerabilities. 
We categorize the vulnerabilities as address-based ($A$) if they are able to derive the complete address of $u$ by 
observing cache hit of $u$ 
and obtaining different timing compared with other candidate data.
Set-based ($S$) vulnerabilities are the ones that can know the mapped set of $u$ by conflicting and generating eviction between $u$ and candidate data addresses.
The third type are the ones that 
potentially derive information from set or address ($SA$) depending
on timing differences derived for all the candidates of $u$. 
For example, $SA$ type
\#33 vulnerability $V_u \rightsquigarrow V_a \rightsquigarrow V_u$ can be set-based if 
$a$ and $u$ are not the same but map to the same cache set, which differs in timing between \{2\}\{8\}\{24\}\{30\}, a local L2 hit, and \{1\}\{7\}\{23\}\{29\}, a local L1 hit. 
Or it can be address-based if $a$ maps to $u$ and $Step\;1$ ($V_u$) and $Step\;2$ ($V_a$) are accessed by different operations (read or write), which have different timing between reads of L1 clean data and dirty data, \{1\} and 
\{7\},
or writes of L1 clean data and dirty data, \{23\} and 
\{29\}.

%% file: micro_benchmark.tex
\section{Benchmark Implementation}
\label{sec:micro_benchmark}

This section presents details of how the benchmarks for the different vulnerabilities are implemented.

\subsection{Benchmarks for Each Vulnerability}
\label{subsec:benchmark_number}
For each vulnerability, there are three steps, and
each step can be two possible cases:
read or write access for a memory access operation;
flush or write in the remote core for an invalidation-related operation.
Thus, there are in total of $2^3=8$ cases considering different types of operations.
Further, if the vulnerabilities have both the victim and the attacker running in one core, these two parties can run either time-slicing or multi-threading. 
Based on that, one case may be doubled for running in two settings.
So for one vulnerability type, there are corresponding 8-16 cases depending on the specific vulnerability.
In total, we found there are 1094 benchmarks for all 88 {\em Strong} type vulnerabilities.

Since each type of states have the same implementation and each three-step benchmark follows a similar pattern, we wrote C scripts to automatically 
generate the C program for each of the 1094 benchmarks.

\iffalse
\subsection{Emulating Victim's Behavior and Attacker's Observations}
\label{subsec:attackers_analysis}

Each of the 88 vulnerabilities contains at least one $V_u$ or $V_u^{inv}$ step, which represents
operation on the secret $u$.
The attacker tries to learn the value of $u$ by guessing if $u$ equals: $a$, $a^{alias}$ or $NIB$.  
$a$ denotes the address that is within the range of sensitive locations $x$ and maps to the target cache line.
$a^{alias}$ denotes any data address that also maps to the cache line but is not the same as $a$.
Furthermore,
it is possible that 
$u$ does not map to the target cache line the attacker is measuring. 
We denote these addresses as ${NIB}$ (not-in-block).
Therefore,
following the states of the three-step model, $u$ can be either $a$, $a^{alias}$, or ${NIB}$.
If the attacker is able to unambiguously correlate the timing to one of the three values, he or she
is able to learn the value of $u$ and the corresponding three-steps is a {\em Strong} type vulnerability.  Meanwhile, if the attacker is not able to clearly distinguish whether
$u$ is $a$, $a^{alias}$, or ${NIB}$ based on the timing, then the corresponding attacks are not ``strongly'' effective (may belong to {\em Weak} type of vulnerabilities if there are still timing difference observed) and may not even be possible.
\fi

\subsection{Judging A Three-Step Combination}

For a specific benchmark that implements one case of the three-step combinations,
following the idea of the cache three-step simulator in Section~\ref{subsec:improve_three},
if the step is
$V_u$, the benchmarks separately test the timing when $V_u$ is $V_a$, $V_{a^{alias}}$, or $V_{NIB}$.
If the step is
$V_u^{inv}$, the benchmarks separately test the timing when $V_u^{inv}$ is $V_a^{inv}$, $V_{a^{alias}}^{inv}$, or $V_{NIB}^{inv}$.
The timing of the last step in the three-step pattern is measured.
For each of the cases, there is $RUN\_NUM$ number of trials,
and Welch's t-test~\cite{welch1947generalization} is used to distinguish the distributions of the measured timings.
We consider two distributions to be significantly different from each other if the probability of observing the data given that they come from the same distribution is less than 0.05\%.

For an effective vulnerability, one of the three candidates of $V_u$ (or $V_u^{inv}$)
should generate timing distribution that is statistically different from the other two candidates.
This is for the {\em Strong} vulnerability types which are 88 types in total. 
The 80 {\em Weak} vulnerability types are not currently considered in the benchmarks
but can be straightforward to add if needed.
At end of each benchmark run, the benchmark outputs if there was significant timing difference--``vulnerability is found'',
or not--``vulnerability not found''.
Further, if any one of the cases of the vulnerability returns ``vulnerability is found'',
the processor cache is considered to be vulnerable to that attack type.

\subsection{Timing Measurement}
\label{subsec:noise}

We use {\em rdtsc} instruction in our benchmarks to do timing measurements, which is the most effective method compared with 
hardware performance counters, which may be limited~\cite{gruss2016flush+} or lacking-determinism~\cite{das2019sok}, or using a ``counting'' thread.
AMD's {\em rdtsc} instruction is not as accurate as 
Intel machine's, but there are plenty of works~\cite{aciiccmez2006trace, irazoqui2016cross, hund2013practical} 
showing that it is also able to be used for cache timing-based attacks.

Noise and variation in the timing measurements could further result in false negatives (if the time measurement was not accurate enough to 
distinguish different timings of  accesses) or false positives (if timing changes resulted in timing measurement differences even though there is no timing difference).

To reduce the noise, instead of measuring just one cache block, we arbitrarily chose 8 cache blocks from different cache sets to
do operation on.
Further, the measurements are all repeated $RUN\_NUM$ times and collect statistical data.
{\em fence} instructions are added between each memory-related instruction to 
enforce an ordering constraint for the attacks.

To reduce the variation of the timing among different cache sets, the timing measurement of the last step is repeated
for each test if the last step is $u$-related step.  Specifically, right after the third step's timing measurement, we trigger and measure the timing of this step again, which is guaranteed to result in an L1 cache hit timing
or timing to invalidate the data that is not in the caches, depending on the concrete memory operations.
We then compare the timing of the third step with the repeated third step. This eliminates any variations in timing among different cache sets.

\iffalse
In this case, if the attacker has better timer source than using {\em rdtsc} to measure timing or using other optimization ways to reduce noise, the attacker may be able to exploit more vulnerabilities than the benchmark.
However, in terms of exploring all possible effective vulnerabilities, when running all 4913 types of vulnerabilities using {\em rdtsc} to measure timing and deploying current settings to reduce noise,
we do not find any other effective vulnerabilities that are not included in our derived effective types as well.
\fi

\begin{figure}[tp]
\centerline{\includegraphics[width=9cm]{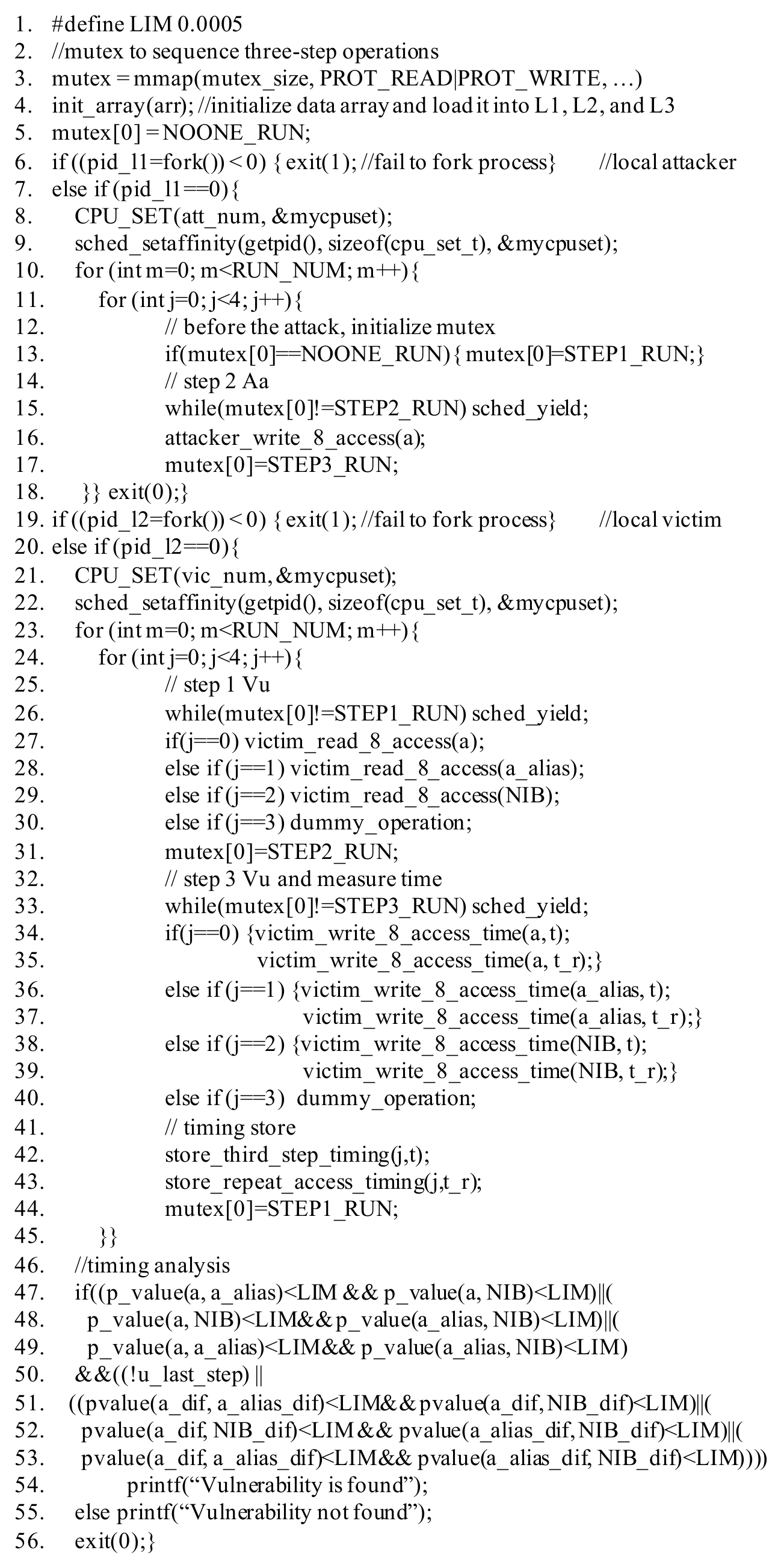}}
\caption{\small The pesudo code of \#42 vulnerability $V_u \rightsquigarrow A_a \rightsquigarrow V_u$ for read ($V_u$), write ($A_a$), and write ($V_u$) case running hyper-threading.}
\label{fig:0017}
\end{figure}

\subsection{Benchmark Code Example}

Figure~\ref{fig:0017} shows an example pseudo code of \#42 vulnerability $V_u \rightsquigarrow A_a \rightsquigarrow V_u$'s benchmark for read ($V_u$), write ($A_a$), and write ($V_u$) access of
the three steps and running hyper-threading.

First, 
we define probability bound of Welch's t-test (line 1) and initialize
a shared array (line 2-3) 
used by mutexes to control the sequence of the three-step accesses.
Then, the 
data (stored in the array) that will be accessed by the victim and the attacker 
is loaded into the L1 cache (line 4), and consequently possibly brought into L2 and L3 caches. 
We use {\em fork()} (line 6 and line 19) to create sub-process, one for the victim and one for the attacker in this example.
Each remote and local victim and attacker will have one sub-process throughout the whole test.
Each sub-process is assigned to a hardware thread (line 8-9, line 21-22).
When running hyper-threading, two local or two remote sub-processes are run in different hardware threads of one CPU, if applicable.
If running time-slicing, sub-processes are assigned to one hardware thread.
Within each sub-process, the test will be run for a certain predefined $RUN\_NUM$ (line 10 and line 23)  times so the timing statistics can be done
based on a large number of runs.
We set $RUN\_NUM$ at 600 to minimize noise and maintain a suitable test set number for Welch's t-test to measure distributions.

As discussed in Section~\ref{subsec:attackers_analysis}, for all the effective vulnerabilities, there will be 
at least one $V_u$ step (or $V_u^{inv}$). 
Within each running of the test, the three values (i.e., $V_a$, $V_{a^{alias}}$, or $V_{NIB}$) will be tested for the $V_u$ or $V_u^{inv}$ (line 11 and line 24).
The ``dummy operation'' branch is used to avoid making the third branch to be the last branch, which we found experimentally has an abnormal timing measurement result.

This particular test, shown in Figure~\ref{fig:0017}, performs $Step\;1$ ($V_u$, line 25-31), $Step\;2$ ($A_a$, line 14-17) and $Step\;3$ ($V_u$, line 32-44).
Last step $Step\;3$ is performed twice and results are stored (line 41-43).
The first access of $Step\;3$ is done to measure if the attacker can observe timing differences when running different values of $u$.
The second access of the third step will always be a hit (fast timing,
and is used to obtain baseline ``fast'' timing for that cache set, as we observed different cache sets can have different timing).
In this case, we can collect results of difference between the first access timing and the second access timing for each candidate of $u$ to 
limit the possibilities that timing difference is due to different cache sets but not different values of $u$.

In the end, Welch's t-test is first applied to each statistical distribution of candidate values for $V_u$ 
(or $V_u^{inv}$)
to see whether the attacker can observe different timing when $V_u$ refers to different addresses (line 47-49). 
If the three-step patterns have $u$-related step as the last step (implemented by {\em u\_last\_step} in line 50), to remove the noise in the timing among different cache sets, the second access timing is considered. 
Welch's t-test is applied to test the difference of the first and the second access of the last step $Step\;3$.
Only if one candidate's distribution has significant timing difference compared with the other two, the cache sets' noise is shown to be not the reason of timing difference and the corresponding vulnerability is judged to be effective (line 51-53).

%% file: security_discussion.tex
\section{Evaluation and Security Discussion}
\label{sec:security_discussion}

\begin{figure*}[htp]
\centerline{\includegraphics[width=18cm]{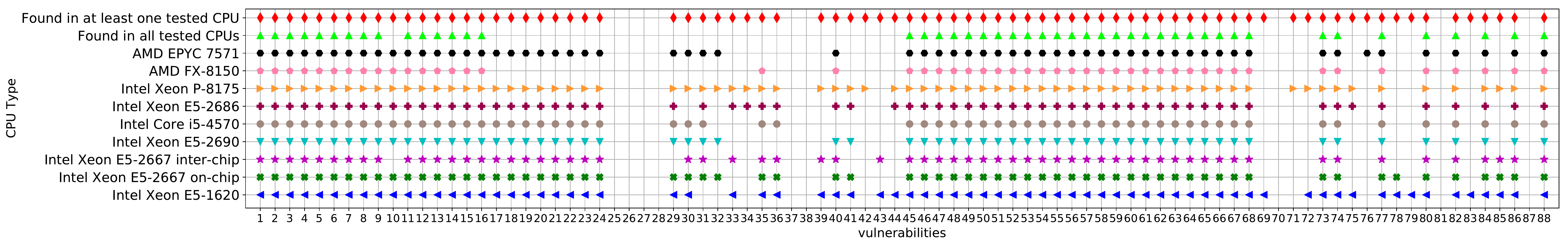}}
\caption{\small Evaluation of 88 {\em Strong} types of vulnerabilities on different machines.
A dot means the corresponding processor is vulnerable to the vulnerability type.
Intel Xeon E5-2667 in our lab has two sockets. 
Therefore, the local and remote core can be both in one socket, i.e., run on-chip; or local and remote core can be in different sockets, i.e., run inter-chip.}
\label{fig:vul_by_type}
\end{figure*}

\begin{figure*}
  \centering
  \begin{tabular}{@{}c@{}}
    \includegraphics[width=18cm]{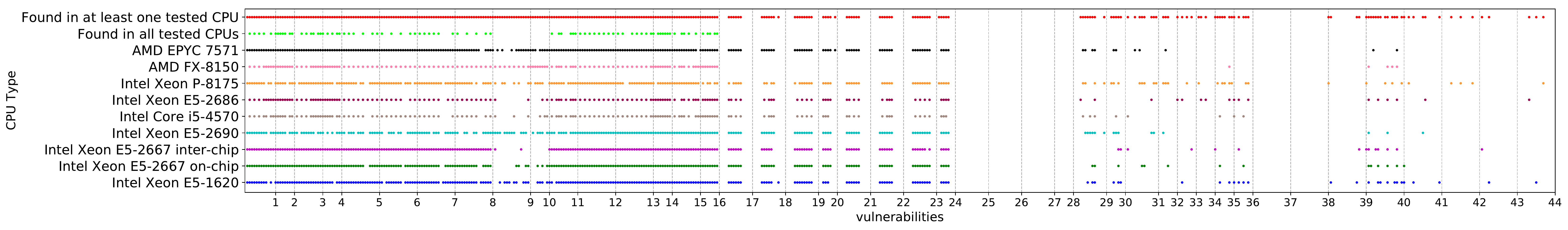} \\[\abovecaptionskip]
    \small (a) \#1 - \#44 vulnerability testing results on different machines. 
  \end{tabular}

  \vspace{0pt} 

  \begin{tabular}{@{}c@{}}
    \includegraphics[width=18cm]{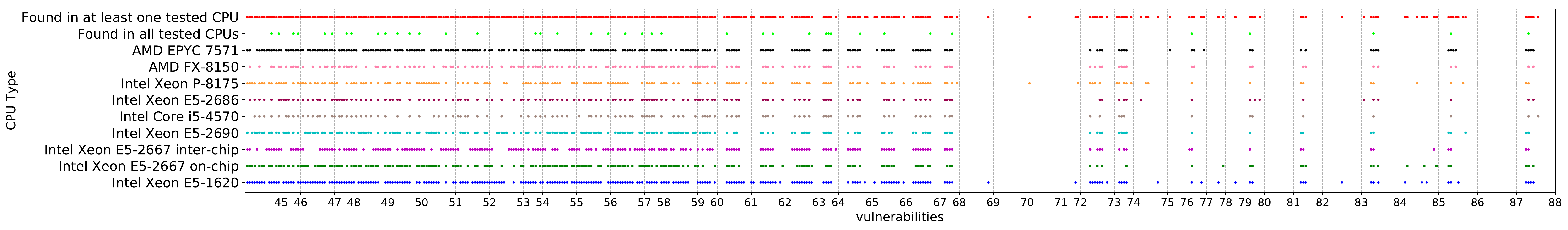} \\[\abovecaptionskip]
    \small (b) \#45 - \#88 vulnerability testing results on different machines. 
  \end{tabular}

  \caption{\small Evaluation of 88 {\em Strong} types of vulnerabilities for all the cases. 
  A dot means the corresponding processor is vulnerable to the vulnerability case.
  For each vulnerability, a fixed number of cases (see Section~\ref{subsec:benchmark_number}) are tested according to the vulnerability type.
  And there are in total 1094 cases for 88 {\em Strong} types of vulnerabilities.
  }\label{fig:vul_by_case}
\end{figure*}

\begin{table}[btp]
\begin{center}
\label{tbl:machine_type}
\caption{\small Configurations of the experimental machines, which all have 64B L1 cache line size. 
(1) denotes the number of hardware threads sharing one L1 cache; (2) denotes the number of hardware threads per socket; (3) denotes the number of sockets.}
 {\fontsize{7}{9}\selectfont
   \begin{tabular}{|C{1.179cm}|C{0.68cm}|C{0.68cm}|C{0.78cm}|C{0.78cm}|C{0.325cm}|C{0.35cm}|C{0.325cm}| }
   \hline

   Model Name & L1-D Cache & L1-I Cache & L2 Cache & L3 Cache & (1) & (2) & (3)  \\ \hline\hline

Intel Xeon E5-1620	&32KB, 8-way	&32KB, 8-way		&256KB, 8-way		&10MB, 20-way		&2	&8	&1	\\ \hline
Intel Xeon E5-2667	&32KB, 8-way	&32KB, 8-way		&256KB, 8-way		&15MB, 20-way		&2	&12	&2	\\ \hline
Intel Xeon E5-2690	&32KB, 8-way	&32KB, 8-way		&256KB, 8-way		&20MB, 20-way		&2	&16	&1	\\ \hline
Intel Core i5-4570	&32KB, 8-way	&32KB, 8-way		&256KB, 8-way		&6MB, 12-way		&1	&4	&1	\\ \hline
Intel Xeon E5-2686	&32KB, 8-way	&32KB, 8-way		&1MB, 16-way		&33MB, 11-way		&1	&4	&1	\\ \hline
Intel Xeon P-8175	&32KB, 8-way	&32KB, 8-way		&256KB, 8-way		&45MB, 20-way		&2	&8	&1	\\ \hline
AMD FX-8150		&16KB, 4-way	&64KB, 2-way		&2MB, 16-way		&8MB, 64-way		&1	&8	&1	\\ \hline
AMD EPYC 7571		&32KB, 8-way	&64KB, 4-way		&512KB		&8MB		&2	&4	&1	\\ \hline

   \end{tabular}
   }
   \label{tbl:machine_type}
 %\vspace{-10pt}
 \end{center}
 \end{table}
 %\end{savenotes}

\textbf{Experimental Setup}. The experimental results reported for Intel processors were performed on 
Intel Core i5-4570, Xeon E5-2690, E5-2667, E5-1620, P-8175 and E5-2686 CPUs. The AMD tests were on AMD EPYC 7571 and AMD FX-8150.
P-8175, E5-2686 and  AMD EPYC 7571 instance are from Amazon EC2.
Table~\ref{tbl:machine_type} shows the processor configurations.

\subsection{Vulnerability Evaluation on Commodity CPUs}

We evaluated 
88 {\em Strong} effective vulnerabilities 
shown in Table~\ref{tbl:attack_list_all}.
Figure~\ref{fig:vul_by_type} lists the results of running the experiments when testing the 9 types of processor configurations upon 88 effective vulnerabilities.
For each type of processors, a dot showing up in the figure
means that the machine is vulnerable to this vulnerability.
Apart from the 9 types of tested processors, 
the first line of dots in Figure~\ref{fig:vul_by_type}
 show the {\em or} result
of 9 types of processor configurations. 
The dot means that the vulnerability is found in at least one tested processor. 
The second line of dots 
in Figure~\ref{fig:vul_by_type} show the {\em and} result
of 9 types of processor configurations. 
The dot means that the vulnerability is found in all tested processors. 

Figure~\ref{fig:vul_by_type} shows that 88 effective vulnerabilities are mostly found in all the tested CPUs. 
Since our new cache three-step simulator considers the ideal case where 66 types of timing observations all have unique results,
it outputs all the possible vulnerability types.
For commodity processors, a subset of them are 
shown to be effective.
This is due to the actual cache
implementation and timing measurement methods, and thus,
some of the timing of the 66 types are not differentiable, as
can be seen from histograms in Figure~\ref{fig:hist}.
Figure~\ref{fig:vul_by_type} also demonstrates that different machines are vulnerable to different types of attacks. 
The {\em and} result
of 9 types of processor configuration experiments have relatively small percentage of vulnerabilities to which machines are all vulnerable.
We further list the statistical results as CTVS for each machine in Section~\ref{sec:ctvs}.

\subsection{Analysis of Vulnerabilities Found}
\label{subsubsec:vul_different_types}

Figure~\ref{fig:vul_by_case} shows the results of 
benchmarks for the 88 vulnerability types.
Machines not supporting hyper-threading have much fewer effective cases.
Similar to Figure~\ref{fig:vul_by_type}, the dot means the related processor is vulnerable to the specific case. 
The gray vertical lines 
are used to group all the cases per vulnerability (there are thus 88 vertical bars and groupings).
We further collect the data in Figure~\ref{fig:vul_by_case} and group them with different $Step\;3$ 
types as the timing observation steps 
in Table~\ref{tbl:success_obser} to compare effects of different operations on cache timing attacks.

 \begin{table}[btp]
\begin{center}
\label{tbl:success_obser}
\caption{\small Percentage of vulnerability cases that are effective for different types of timing observation steps for different machine configurations. The number on the right of ``/'' is the total cases of vulnerabilities for the corresponding categorization; the number on the left of ``/'' is the number of cases to which the corresponding processor is vulnerable.
Machines labeled * do not support hyper-threading.}
 {\fontsize{7}{9}\selectfont
   \begin{tabular}{|C{2.15cm}|C{0.97cm}|C{0.97cm}|C{1.22cm}|C{1.05cm}| }
   \hline

  Model Name  & Local Read & Local Write & Remote Write to Inv. & Flush to Inv. \\ \hline\hline

Intel Xeon E5-1620				&137/277	&118/277	&127/277	&129/263	\\ \hline
Intel Xeon E5-2667 on-chip		&121/277	&117/277	&80/277		&119/263	\\ \hline
Intel Xeon E5-2667 inter-chip	&127/277	&111/277	&124/277	&72/263	\\ \hline
Intel Xeon E5-2690				&128/277	&101/277	&77/277		&107/263	\\ \hline
Intel Core i5-4570*				&82/277		&66/277		&57/277		&63/263	\\ \hline
Intel Xeon E5-2686*				&87/277		&74/277		&69/277		&80/263	\\ \hline
Intel Xeon P-8175				&124/277	&120/277	&75/277		&105/263	\\ \hline
AMD FX-8150*					&68/277		&65/277		&89/277		&65/263	\\ \hline
AMD EPYC 7571 					&125/277	&125/277	&124/277	&114/263	\\ \hline

in all CPUs						&49/277		&34/277		&10/277		&30/263	\\ \hline
at least one CPU				&175/277	&150/277	&162/277	&162/263	\\ \hline
   \end{tabular}
   }
   \label{tbl:success_obser}
 \end{center}
 \end{table}

\textbf{Local read and local write of timing
 observation step.}
\label{subsec:read_write}
Previous attacks normally used {\em read} access to implement the side-channel attacks, as analyzed in Section~\ref{subsubsec:read_write}.
However, write access is shown in Figure~\ref{fig:vul_by_case} and Table~\ref{tbl:success_obser} to be an effective 
method to implement attacks as well.
It has generally smaller rate
compared with read access to trigger effective vulnerabilities of different cases, especially for tested machine Intel Xeon E5-1620 and E5-2690. 
For the 44 types of vulnerabilities (\#1 - \#44) that have access operation as timing observation step, 
Figure~\ref{fig:vul_by_case} demonstrates that
there are 38 out of 44 vulnerabilities to which at least one machine is vulnerable when using read as the timing observation step.
While using write access as the timing observation step, 
34 out of 44 vulnerabilities are vulnerable to at least one machine.

\textbf{Invalidation using cache coherence or flush for timing
 observation step.}
According to Table~\ref{tbl:success_obser}, the percentage of vulnerabilities 
to which the machine is vulnerable
mainly depends on processor types 
when comparing different invalidation-related operation as the timing observation step.
Among the tested processors, Intel Xeon E5-2667 running inter-chip, AMD FX-8150 and AMD EPYC are more vulnerable to remote write as the timing observation step.
 Intel Xeon E5-1620, E5-2667 running on-chip, E5-2690, E5-2686, P-8175, and Core i5-4570 are more vulnerable to flush observation step. 
Overall, for the 44 types of vulnerabilities (\#45 - \#88) that have invalidation for timing observation, remote write and flush operations both have
38 out of 44 vulnerabilities to which at least one machine is vulnerable.

\begin{table}[btp]
\begin{center}
\label{tbl:success_th}
\caption{\small Percentage of vulnerability cases that are effective for 
the victim (Vic.) and the attacker (Att.) running the same core (time-slicing or hyper-threading), 
running different cores or within the victim
for different machine configurations.
The number on the right of ``/'' is the total cases of vulnerabilities for the corresponding categorization; the number on the left of ``/'' is the number of cases to which the corresponding processor is vulnerable.
Machines labeled * do not support hyper-threading.}
 {\fontsize{7}{9}\selectfont
   \begin{tabular}{|C{2.08cm}|C{1.12cm}|C{1.12cm}|C{1.28cm}|C{0.85cm}| }%C{3cm}|C{2cm}|C{2cm}| }
   \hline

   \multirow{3}{*}{\shortstack{Model Name}} & \multicolumn{2}{ c |}{ Vic., Att. on the Same Core} &  \multirow{3}{*}{\shortstack{Vic., Att. on\\ Different\\ Cores}} & \multirow{3}{*}{\shortstack{Within \\Victim}}  \\ \cline{2-3}
    ~ & Time-Slicing & Hyper-Threading & ~ & ~ \\ \hline\hline

Intel Xeon E5-1620				&181/390	&174/390	&51/90		&105/224	\\ \hline
Intel Xeon E5-2667 on-chip		&156/390	&146/390	&52/90		&83/224	\\ \hline
Intel Xeon E5-2667 inter-chip	&151/390	&146/390	&49/90		&88/224	\\ \hline
Intel Xeon E5-2690				&144/390	&138/390	&46/90		&85/224	\\ \hline
Intel Core i5-4570*				&143/390	&0/390		&46/90		&79/224	\\ \hline
Intel Xeon E5-2686*				&166/390	&0/390		&50/90		&94/224	\\ \hline
Intel Xeon P-8175				&148/390	&143/390	&38/90		&95/224	\\ \hline
AMD FX-8150*					&155/390	&0/390		&43/90		&89/224	\\ \hline
AMD EPYC 7571 					&159/390	&171/390	&55/90		&103/224	\\ \hline

in all CPUs						&67/390		&0/390		&18/90		&38/224	\\ \hline
at least one CPU				&223/390	&217/390	&61/90		&148/224	\\ \hline

   \end{tabular}
   }
   \label{tbl:success_th}
 \end{center}
 \end{table}

\textbf{Running time-slicing or hyper-threading.}
Besides different kinds of operations, we also collect results in Table~\ref{tbl:success_th} for running time-slicing and 
 hyper-threading  when the victim and the attacker run on the same core (either local core or remote core).
 There are also vulnerabilities for which the victim and the attacker run on different cores, 
 or vulnerabilities only having victim steps.
 Based on the results, running time-slicing 
is more vulnerable compared with running hyper-threading for Intel processors.
While AMD processor
EPYC 7571 shows that running time-slicing is more vulnerable.
 Furthermore, hyper-threading provides more choices for the attacker to exploit the corresponding vulnerability in different ways.

%% file: ctvs.tex
\section{Cache Timing Vulnerability Score}
\label{sec:ctvs}

\begin{table}[t]
\begin{center}
\label{tbl:ctvs}
\caption{\small Cache Timing Vulnerability Score (CTVS) for each of the tested processors. 
The number on the right of ``/'' is the total cases of vulnerabilities for the corresponding categorization; the number on the left of ``/'' is the number of types to which the corresponding processor is vulnerable.
Smaller is better.
``$E$'' and ``$I$'' are internal and external interference vulnerabilities, respectively.
 ``$S$'' and ``$A$'' are set-based and address-based vulnerabilities, respectively.
``$SA$'' are the ones that are both set-based and address-based.}
 {\fontsize{7}{9}\selectfont
   \begin{tabular}{|C{1.077cm}|C{0.675cm}|C{0.597cm}|C{0.597cm}|C{0.597cm}|C{0.597cm}|C{0.597cm}|C{0.597cm}| }
   \hline

   Model Name & CTVS Score & $I$-$A$ Vul. & $I$-$S$ Vul. & $I$-$SA$ Vul. & $E
   $-$A$ Vul. & $E$-$S$ Vul. & $E$-$SA$ Vul. \\ \hline\hline

Intel Xeon E5-1620				&73/88	&20/20	&6/12	&12/12	&20/20	&5/12	&10/12	\\ \hline
Intel Xeon E5-2667 on-chip		&66/88	&20/20	&3/12	&11/12	&20/20	&3/12	&9/12	\\ \hline
Intel Xeon E5-2667 inter-chip	&64/88	&19/20	&2/12	&12/12	&20/20	&3/12	&8/12	\\ \hline
Intel Xeon E5-2690				&62/88	&20/20	&1/12	&10/12	&20/20	&2/12	&9/12	\\ \hline
Intel Core i5-4570	 			&61/88	&20/20	&1/12	&10/12	&20/20	&1/12	&9/12	\\ \hline
Intel Xeon E5-2686				&66/88	&20/20	&2/12	&11/12	&20/20	&3/12	&10/12	\\ \hline
Intel Xeon P-8175				&73/88	&20/20	&5/12	&12/12	&20/20	&5/12	&11/12	\\ \hline
AMD FX-8150						&50/88	&18/20	&1/12	&7/12	&18/20	&0/12	&6/12	\\ \hline
AMD EPYC 7571 					&62/88	&20/20	&2/12	&10/12	&20/20	&1/12	&9/12	\\ \hline

in all CPUs						&47/88	&17/20	&0/12	&6/12	&18/20	&0/12	&6/12	\\ \hline
at least one CPU				&79/88	&20/20	&9/12	&12/12	&20/20	&7/12	&11/12	\\ \hline

   \end{tabular}
   }
   \label{tbl:ctvs}
\end{center}
\end{table}

Table~\ref{tbl:ctvs} shows the CTVS values for each tested machine,
as well as the {\em and} result and {\em or} result for machines.
For CTVS number, smaller is better.
For all the 88 {\em Strong} type vulnerabilities,
AMD FX-8150 has relatively better CTVS compared with Intel machines.
Xeon E5-1620 and P-8175 are the most vulnerable ones among Intel processors.
Otherwise, the Xeon family and AMD EPYC 7571 have generally similar results.

In Table~\ref{tbl:ctvs},
CTVS numbers for the {\em and} result of all processors are small. This
demonstrates that
only a few attacks can be effective in all the processors.
These numbers are expected to be even smaller if more processors are tested.
CTVS numbers for the {\em or} result of all processors are large.
This
confirms that nearly all of the vulnerabilities derived by the new three-step model are found in the tested processors.

$A$ type vulnerabilities generally have higher effective rates than $S$ type vulnerabilities.
This is because that $S$ type vulnerabilities
normally 
differentiate timing between L1 cache and L2 cache, i.e., accessing or invalidating L1 or L2 data,
which are shown in the histograms in Figure~\ref{fig:hist} to be much smaller compared with the difference between L1 cache hit and DRAM hit, for example.
Especially, the timing difference between 
remote write to invalidate dirty L1 data and L2 data is almost non-differentiable,
resulting in that related vulnerabilities (especially \#25 - \#28) are found to be 
not effective in all tested processors (shown in Figure~\ref{fig:vul_by_type}).
$A$ type vulnerabilities generally rely on timing differences between
L1 cache hit and DRAM hit, or L1 cache hit and remote L1 cache hit;
histograms in Figure~\ref{fig:hist} demonstrate that these access types have large timing differences,
making these vulnerabilities much more effective.
$SA$ type vulnerabilities generally leverage
the timing differences between clean L1 data invalidation and dirty L1 data invalidation 
or between local access of remote clean L1 data and remote dirty L1 data;
histograms in Figure~\ref{fig:hist} again show large timing differences for these, and 
related vulnerabilities are found to be very effective by CTVS.

For $I$ and $E$ type vulnerabilities,
they do not have an explicit distinction of CTVS numbers for the tested processors.
\iffalse
the $I$ type vulnerabilities are more common on the processors, 
which may be result of smaller noise because of fewer process switches during execution of the benchmarks.
\textbf{FIXME is this correct?  "process switching" is never mentioned when discussing design of benchmarks}
For a specific vulnerabilities, if process switch is needed \textbf{FIXME again, did not clearly say anything about process switches when discussing benchmark design} between different steps, the number of process switches are added up, 
\textbf{FIXME they are added up to what?  more process switches equals more vulnerabilities? Figure~\ref{fig:0017} does not mention process switches,
nor can you see any "adding up" of process switches in Figure~\ref{fig:0017}}
e.g., \#42 vulnerability $V_u \rightsquigarrow A_a \rightsquigarrow V_u$ shown in Figure~\ref{fig:0017} has two process switches.
$I$ type vulnerabilities therefore have fewer process switches and less noise than $E$ type vulnerabilities.
This especially applies for machines such as AMD EPYC 7571.
\textbf{FIXME unclear, if I has fewer switches than E, then I should be better on all machines, not just AMD, number of switches does not depend on Intel vs. AMD.}
\fi

\subsection{for Design of Defense Features}

CTVS can be used to test existing processors, as we have shown in this work.
Further, the benchmarks can be used to evaluate any new secure cache designs,
by running the benchmarks on a simulator.

CTVS has shown that different processors are vulnerable to different attacks.
Consequently, customized software or hardware defenses can be deployed for each processor based on the CTVS score,
rather than defending
vulnerabilities not present in the specific processor's caches.  
For software defenses, the access patterns from the benchmarks
could be used as a reference for scanning software to find if it has similar patterns, 
e.g., to find malicious software that has such attack patterns.  Further, CTVS and our three-step model have shown new attack types which are unknown before,
and thus not considered by defenses based on monitoring performance counters --
this points to the requirement of using new or different 
performance counter types in works that use monitoring.

The CTVS can also be used to help understand the implementation of different processors especially the micro-architectures.
In particular, processors that have similar implementations may have similar results of CTVS numbers,
based on common architectural features implemented for the whole memory hierarchy.

%% file: validation.tex
\section{Validation}
\label{sec:validation}

To validate if there
are any other vulnerabilities that are left out apart from all the effective 
vulnerabilities we derived from our cache three-step simulator,
we empirically ran benchmarks for all the cases of the whole $17^3=4913$ three-step combinations for 9 processor configurations.

We discovered a number of three-steps,
besides the {\em Strong}, {\em Weak} and repeat types, returned by the benchmarks to have timing variations but consider
all of them as false positives.
The false positives that show up in every processor we tested all have the second or the third step to be $A^{inv}$, $V^{inv}$ or $\star$.
The corresponding types cannot be any effective vulnerabilities 
because these three types of states will make the attacker lose track of useful information due to whole cache flush ($A^{inv}$, $V^{inv}$)
or zero-knowledge state inference ($\star$) if they are in $Step\;2$ or $Step\;3$.
Reason of three-steps with the second or the third step as $A^{inv}$, $V^{inv}$ to seem to be 
effective in the result of running the benchmarks is that whole cache flush currently cannot be implemented under user-level privilege. 
We use approximate method to implement these states in the benchmark by invalidating every address that is related to the attacks.
An approximate method is also used for $\star$ to simulate the zero-knowledge state.
Therefore, the timings of $A^{inv}$, $V^{inv}$ and $\star$ are not accurate to measure for the real attack.

Overall, we found that there are no effective vulnerabilities that are not covered by the vulnerabilities we derived.
Detailed analysis is not shown due to space limits of the paper.

%% file: conclusion.tex
\section{Conclusion}
\label{sec:conclusion}

This work presented a new three-step model and the first benchmark suite for evaluating all 88 possible {\em Strong} cache timing-based attack types in processors.
The model allowed us to find 32 new timing attack types.
Further, we implemented scripts to auto-generate the 1094 benchmark tests  from our three-step model's 88 theoretical attack types for testing different cases.
The benchmarks were run on
a number of commodity processors
to gave each machine the Cache Timing Vulnerability Score (CTVS) to measure the degree of the machine's robustness against cache timing-based vulnerabilities.
The three-step model, benchmarks, and the CTVS can be used to measure existing systems and help design future secure caches.

All code related to this work will be made open-source.

%% file: ms.bbl
\begin{thebibliography}{10}

\bibitem{aciiccmez2006trace}
Onur Ac{\i}i{\c{c}}mez and {\c{C}}etin~Kaya Ko{\c{c}}.
\newblock {Trace-Driven Cache Attacks on AES (short paper)}.
\newblock In {\em International Conference on Information and Communications
  Security}, pages 112--121. Springer, 2006.

\bibitem{bernstein2005cache}
Daniel~J Bernstein.
\newblock {Cache-Timing Attacks on AES}.
\newblock 2005.

\bibitem{bonneau2006cache}
Joseph Bonneau and Ilya Mironov.
\newblock {Cache-Collision Timing Attacks against AES}.
\newblock In {\em International Workshop on Cryptographic Hardware and Embedded
  Systems}, pages 201--215. Springer, 2006.

\bibitem{bourgeat2018mi6}
Thomas Bourgeat, Ilia Lebedev, Andrew Wright, Sizhuo Zhang, Arvind, and
  Srinivas Devadas.
\newblock {MI6: Secure Enclaves in a Speculative Out-of-Order Processor}.
\newblock {\em arXiv preprint arXiv:1812.09822}, 2018.

\bibitem{storebypass2018}
INTEL CORP.
\newblock {Speculative Store Bypass Bug CVE, 2018. CVE 2018-3639.}
\newblock \url{https://cve.mitre.org/ cgi-bin/cvename.cgi?name=CVE-2018-3639},
  accessed online.

\bibitem{costan2016sanctum}
Victor Costan, Ilia~A Lebedev, and Srinivas Devadas.
\newblock {Sanctum: Minimal Hardware Extensions for Strong Software Isolation}.
\newblock In {\em USENIX Security Symposium}, pages 857--874, 2016.

\bibitem{crane2015thwarting}
Stephen Crane, Andrei Homescu, Stefan Brunthaler, Per Larsen, and Michael
  Franz.
\newblock {Thwarting Cache Side-Channel Attacks Through Dynamic Software
  Diversity}.
\newblock In {\em NDSS}, pages 8--11, 2015.

\bibitem{daemen1999aes}
Joan Daemen and Vincent Rijmen.
\newblock {AES Proposal: Rijndael}.
\newblock 1999.

\bibitem{das2019sok}
Sanjeev Das, Jan Werner, Manos Antonakakis, Michalis Polychronakis, and Fabian
  Monrose.
\newblock {SoK: The Challenges, Pitfalls, and Perils of Using Hardware
  Performance Counters for Security}.
\newblock In {\em Proceedings of 40th IEEE Symposium on Security and Privacy
  (S\&P’19)}, 2019.

\bibitem{demme2012side}
John Demme, Robert Martin, Adam Waksman, and Simha Sethumadhavan.
\newblock {Side-Channel Vulnerability Factor: A Metric for Measuring
  Information Leakage}.
\newblock In {\em 2012 39th Annual International Symposium on Computer
  Architecture (ISCA)}, pages 106--117. IEEE, 2012.

\bibitem{deng2019analysis}
Shuwen Deng, Wenjie Xiong, and Jakub Szefer.
\newblock {Analysis of Secure Caches using a Three-Step Model for Timing-Based
  Attacks}.
\newblock Cryptology ePrint Archive, Report 2019/167, 2019.
\newblock \url{https://eprint.iacr.org/2019/167}.

\bibitem{Deng:2019:ST:3307650.3322238}
Shuwen Deng, Wenjie Xiong, and Jakub Szefer.
\newblock {Secure TLBs}.
\newblock In {\em Proceedings of the 46th International Symposium on Computer
  Architecture}, ISCA '19, pages 346--259, New York, NY, USA, 2019. ACM.

\bibitem{domnitser2010predictive}
Leonid Domnitser, Nael Abu-Ghazaleh, and Dmitry Ponomarev.
\newblock {A Predictive Model for Cache-Based Side Channels in Multicore and
  Multithreaded Microprocessors}.
\newblock In {\em International Conference on Mathematical Methods, Models, and
  Architectures for Computer Network Security}, pages 70--85. Springer, 2010.

\bibitem{domnitser2012non}
Leonid Domnitser, Aamer Jaleel, Jason Loew, Nael Abu-Ghazaleh, and Dmitry
  Ponomarev.
\newblock {Non-Monopolizable Caches: Low-Complexity Mitigation of Cache Side
  Channel Attacks}.
\newblock {\em ACM Transactions on Architecture and Code Optimization (TACO)},
  8(4):35, 2012.

\bibitem{gras2018translation}
Ben Gras, Kaveh Razavi, Herbert Bos, and Cristiano Giuffrida.
\newblock {Translation Leak-aside Buffer: Defeating Cache Side-channel
  Protections with TLB Attacks}.
\newblock In {\em 27th $\{$USENIX$\}$ Security Symposium ($\{$USENIX$\}$
  Security 18)}, pages 955--972, 2018.

\bibitem{gruss2016flush+}
Daniel Gruss, Cl{\'e}mentine Maurice, Klaus Wagner, and Stefan Mangard.
\newblock {Flush+ Flush: a Fast and Stealthy Cache Attack}.
\newblock In {\em International Conference on Detection of Intrusions and
  Malware, and Vulnerability Assessment}, pages 279--299. Springer, 2016.

\bibitem{gruss2015cache}
Daniel Gruss, Raphael Spreitzer, and Stefan Mangard.
\newblock {Cache Template Attacks: Automating Attacks on Inclusive Last-Level
  Caches}.
\newblock In {\em USENIX Security Symposium}, pages 897--912, 2015.

\bibitem{guanciale2016cache}
Roberto Guanciale, Hamed Nemati, Christoph Baumann, and Mads Dam.
\newblock {Cache Storage Channels: Alias-Driven Attacks and Verified
  Countermeasures}.
\newblock In {\em Security and Privacy (SP), 2016 IEEE Symposium on}, pages
  38--55. IEEE, 2016.

\bibitem{gullasch2011cache}
David Gullasch, Endre Bangerter, and Stephan Krenn.
\newblock {Cache Games--Bringing Access-Based Cache Attacks on AES to
  Practice}.
\newblock In {\em Security and Privacy (SP), 2011 IEEE Symposium on}, pages
  490--505. IEEE, 2011.

\bibitem{he2017secure}
Zecheng He and Ruby~B Lee.
\newblock {How Secure is Your Cache against Side-Channel Attacks?}
\newblock In {\em International Symposium on Microarchitecture (MICRO)}, pages
  341--353. ACM, 2017.

\bibitem{hund2013practical}
Ralf Hund, Carsten Willems, and Thorsten Holz.
\newblock {Practical Timing Side Channel Attacks Against Kernel Space ASLR}.
\newblock In {\em 2013 IEEE Symposium on Security and Privacy}, pages 191--205.
  IEEE, 2013.

\bibitem{irazoqui2016cross}
Gorka Irazoqui, Thomas Eisenbarth, and Berk Sunar.
\newblock {Cross Processor Cache Attacks}.
\newblock In {\em Proceedings of the 11th ACM on Asia conference on computer
  and communications security}, pages 353--364. ACM, 2016.

\bibitem{kayaalp2017ric}
Mehmet Kayaalp, Khaled~N Khasawneh, Hodjat~Asghari Esfeden, Jesse Elwell, Nael
  Abu-Ghazaleh, Dmitry Ponomarev, and Aamer Jaleel.
\newblock {RIC: Relaxed Inclusion Caches for Mitigating LLC Side-Channel
  attacks}.
\newblock In {\em Design Automation Conference (DAC), 2017 54th ACM/EDAC/IEEE},
  pages 1--6. IEEE, 2017.

\bibitem{keramidas2008non}
Georgios Keramidas, Alexandros Antonopoulos, Dimitrios~N Serpanos, and Stefanos
  Kaxiras.
\newblock {Non Deterministic Caches: A Simple and Effective Defense against
  Side Channel Attacks}.
\newblock {\em Design Automation for Embedded Systems}, 12(3):221--230, 2008.

\bibitem{kirianskydawg}
Vladimir Kiriansky, Ilia Lebedev, Saman Amarasinghe, Srinivas Devadas, and Joel
  Emer.
\newblock {DAWG: A Defense Against Cache Timing Attacks in Speculative
  Execution Processors}.

\bibitem{Kocher2018spectre}
Paul Kocher, Daniel Genkin, Daniel Gruss, Werner Haas, Mike Hamburg, Moritz
  Lipp, Stefan Mangard, Thomas Prescher, Michael Schwarz, and Yuval Yarom.
\newblock {Spectre Attacks: Exploiting Speculative Execution}.
\newblock {\em ArXiv e-prints}, January 2018.

\bibitem{kong2009hardware}
Jingfei Kong, Onur Acii{\c{c}}mez, Jean-Pierre Seifert, and Huiyang Zhou.
\newblock {Hardware-Software Integrated Approaches to Defend against Software
  Cache-Based Side Channel Attacks}.
\newblock In {\em 2009 IEEE 15th International Symposium on High Performance
  Computer Architecture}, pages 393--404. IEEE, 2009.

\bibitem{kopf2012automatic}
Boris K{\"o}pf, Laurent Mauborgne, and Mart{\'\i}n Ochoa.
\newblock {Automatic Quantification of Cache Side-Channels}.
\newblock In {\em International Conference on Computer Aided Verification},
  pages 564--580. Springer, 2012.

\bibitem{koruyeh2018spectre}
Esmaeil~Mohammadian Koruyeh, Khaled~N Khasawneh, Chengyu Song, and Nael
  Abu-Ghazaleh.
\newblock {Spectre Returns! Speculation Attacks Using the Return Stack Buffer}.
\newblock In {\em 12th $\{$USENIX$\}$ Workshop on Offensive Technologies
  ($\{$WOOT$\}$ 18)}, 2018.

\bibitem{lee2005architecture}
Ruby~B Lee, Peter Kwan, John~P McGregor, Jeffrey Dwoskin, and Zhenghong Wang.
\newblock {Architecture for Protecting Critical Secrets in Microprocessors}.
\newblock In {\em ACM SIGARCH Computer Architecture News}, volume~33, pages
  2--13. IEEE Computer Society, 2005.

\bibitem{Lipp2018meltdown}
Moritz Lipp, Michael Schwarz, Daniel Gruss, Thomas Prescher, Werner Haas,
  Stefan Mangard, Paul Kocher, Daniel Genkin, Yuval Yarom, and Mike Hamburg.
\newblock {Meltdown}.
\newblock {\em ArXiv e-prints}, January 2018.

\bibitem{liu2016catalyst}
Fangfei Liu, Qian Ge, Yuval Yarom, Frank Mckeen, Carlos Rozas, Gernot Heiser,
  and Ruby~B Lee.
\newblock {CATalyst: Defeating Last-Level Cache Side Channel Attacks in Cloud
  Computing}.
\newblock In {\em High Performance Computer Architecture (HPCA), 2016 IEEE
  International Symposium on}, pages 406--418. IEEE, 2016.

\bibitem{liu2014random}
Fangfei Liu and Ruby~B Lee.
\newblock {Random Fill Cache Architecture}.
\newblock In {\em Microarchitecture (MICRO), 2014 47th Annual IEEE/ACM
  International Symposium on}, pages 203--215. IEEE, 2014.

\bibitem{liu2015last}
Fangfei Liu, Yuval Yarom, Qian Ge, Gernot Heiser, and Ruby~B Lee.
\newblock {Last-Level Cache Side-Channel Attacks are Practical}.
\newblock In {\em 2015 IEEE Symposium on Security and Privacy}, pages 605--622.
  IEEE, 2015.

\bibitem{maurice2015c5}
Cl{\'e}mentine Maurice, Christoph Neumann, Olivier Heen, and Aur{\'e}lien
  Francillon.
\newblock {C5: Cross-Cores Cache Covert Channel}.
\newblock In {\em International Conference on Detection of Intrusions and
  Malware, and Vulnerability Assessment}, pages 46--64. Springer, 2015.

\bibitem{osvik2006cache}
Dag~Arne Osvik, Adi Shamir, and Eran Tromer.
\newblock {Cache Attacks and Countermeasures: the Case of AES}.
\newblock In {\em Cryptographers’ Track at the RSA Conference}, pages 1--20.
  Springer, 2006.

\bibitem{percival2005cache}
Colin Percival.
\newblock {Cache Missing for Fun and Profit}, 2005.

\bibitem{qureshi2018ceaser}
Moinuddin~K Qureshi.
\newblock {CEASER: Mitigating Conflict-Based Cache Attacks via
  Encrypted-Address and Remapping}.
\newblock In {\em 2018 51st Annual IEEE/ACM International Symposium on
  Microarchitecture (MICRO)}, pages 775--787. IEEE, 2018.

\bibitem{sabbagh2018scadet}
Majid Sabbagh, Yunsi Fei, Thomas Wahl, and A~Adam Ding.
\newblock {SCADET: A Side-Channel Attack Detection Tool for Tracking
  Prime-Probe}.
\newblock In {\em 2018 IEEE/ACM International Conference on Computer-Aided
  Design (ICCAD)}, pages 1--8. IEEE, 2018.

\bibitem{schwarz2018netspectre}
Michael Schwarz, Martin Schwarzl, Moritz Lipp, and Daniel Gruss.
\newblock {Netspectre: Read Arbitrary Memory over Network}.
\newblock {\em arXiv preprint arXiv:1807.10535}, 2018.

\bibitem{trippel2018meltdownprime}
Caroline Trippel, Daniel Lustig, and Margaret Martonosi.
\newblock {Meltdownprime and Spectreprime: Automatically-Synthesized Attacks
  Exploiting Invalidation-Based Coherence Protocols}.
\newblock {\em arXiv preprint arXiv:1802.03802}, 2018.

\bibitem{wang2017cached}
Shuai Wang, Pei Wang, Xiao Liu, Danfeng Zhang, and Dinghao Wu.
\newblock {CacheD: Identifying Cache-Based Timing Channels in Production
  Software}.
\newblock In {\em 26th $\{$USENIX$\}$ Security Symposium ($\{$USENIX$\}$
  Security 17)}, pages 235--252, 2017.

\bibitem{wang2016secdcp}
Yao Wang, Andrew Ferraiuolo, Danfeng Zhang, Andrew~C Myers, and G~Edward Suh.
\newblock {SecDCP: Secure Dynamic Cache Partitioning for Efficient Timing
  Channel Protection}.
\newblock In {\em Design Automation Conference (DAC), 2016 53nd ACM/EDAC/IEEE},
  pages 1--6. IEEE, 2016.

\bibitem{wang2007new}
Zhenghong Wang and Ruby~B Lee.
\newblock {New Cache Designs for Thwarting Software Cache-Based Side Channel
  Attacks}.
\newblock In {\em ACM SIGARCH Computer Architecture News}, volume~35, pages
  494--505. ACM, 2007.

\bibitem{wang2008novel}
Zhenghong Wang and Ruby~B Lee.
\newblock {A Novel Cache Architecture with Enhanced Performance and Security}.
\newblock In {\em Microarchitecture, 2008. MICRO-41. 2008 41st IEEE/ACM
  International Symposium on}, pages 83--93. IEEE, 2008.

\bibitem{welch1947generalization}
Bernard~L Welch.
\newblock {The Generalization of Student's Problem When Several Different
  Population Variances are Involved}.
\newblock {\em Biometrika}, 34(1/2):28--35, 1947.

\bibitem{wernerscattercache}
Mario Werner, Thomas Unterluggauer, Lukas Giner, Michael Schwarz, Daniel Gruss,
  and Stefan Mangard.
\newblock {ScatterCache: Thwarting Cache Attacks via Cache Set Randomization}.
\newblock In {\em 28th {USENIX} Security Symposium ({USENIX} Security 19)},
  Santa Clara, CA, 2019. {USENIX} Association.

\bibitem{xiong2019leaking}
Wenjie Xiong and Jakub Szefer.
\newblock {Leaking Information Through Cache LRU States}.
\newblock {\em arXiv preprint arXiv:1905.08348}, 2019.

\bibitem{yan2018invisispec}
Mengjia Yan, Jiho Choi, Dimitrios Skarlatos, Adam Morrison, Christopher
  Fletcher, and Josep Torrellas.
\newblock {InvisiSpec: Making Speculative Execution Invisible in the Cache
  Hierarchy}.
\newblock In {\em 2018 51st Annual IEEE/ACM International Symposium on
  Microarchitecture (MICRO)}, pages 428--441. IEEE, 2018.

\bibitem{yan2017secure}
Mengjia Yan, Bhargava Gopireddy, Thomas Shull, and Josep Torrellas.
\newblock {Secure Hierarchy-Aware Cache Replacement Policy (SHARP): Defending
  Against Cache-Based Side Channel Attacks}.
\newblock In {\em Proceedings of the 44th Annual International Symposium on
  Computer Architecture}, pages 347--360. ACM, 2017.

\bibitem{yan2019attack}
Mengjia Yan, Read Sprabery, Bhargava Gopireddy, Christopher Fletcher, Roy
  Campbell, and Josep Torrellas.
\newblock {Attack Directories, Not Caches: Side Channel Attacks in a
  Non-inclusive World}.
\newblock In {\em Attack Directories, Not Caches: Side Channel Attacks in a
  Non-Inclusive World}, page~0. IEEE, 2019.

\bibitem{yarom2014flush+}
Yuval Yarom and Katrina Falkner.
\newblock {FLUSH+ RELOAD: A High Resolution, Low Noise, L3 Cache Side-Channel
  Attack.}
\newblock In {\em USENIX Security Symposium}, pages 719--732, 2014.

\bibitem{zhang2012language}
Danfeng Zhang, Aslan Askarov, and Andrew~C Myers.
\newblock {Language-Based Control and Mitigation of Timing Channels}.
\newblock {\em ACM SIGPLAN Notices}, 47(6):99--110, 2012.

\bibitem{zhang2015hardware}
Danfeng Zhang, Yao Wang, G~Edward Suh, and Andrew~C Myers.
\newblock {A Hardware Design Language for Timing-Sensitive Information-Flow
  Security}.
\newblock In {\em ACM SIGARCH Computer Architecture News}, volume~43, pages
  503--516. ACM, 2015.

\bibitem{zhang2014new}
Tianwei Zhang and Ruby~B Lee.
\newblock {New Models of Cache Architectures Characterizing Information Leakage
  from Cache Side Channels}.
\newblock In {\em Proceedings of the 30th Annual Computer Security Applications
  Conference}, pages 96--105. ACM, 2014.

\bibitem{zhang2013side}
Tianwei Zhang, Fangfei Liu, Si~Chen, and Ruby~B Lee.
\newblock {Side Channel Vulnerability Metrics: the Promise and the Pitfalls}.
\newblock In {\em Proceedings of the 2nd International Workshop on Hardware and
  Architectural Support for Security and Privacy}, page~2. ACM, 2013.

\end{thebibliography}
